\documentclass[final,1p,times,authoryear]{elsarticle}
\usepackage{amssymb}
\usepackage{amsmath}
\usepackage{url}
\usepackage{siunitx}
\usepackage{multirow}
\usepackage{multicol}
\usepackage[dvipsnames]{xcolor}
\usepackage{caption}
\usepackage{subcaption}
\usepackage{bbm}
\usepackage{csquotes}
\usepackage{booktabs}  % \cmidrule
\usepackage{hyperref}

\DeclareMathOperator{\swiglu}{SwiGLU}

\DeclareMathOperator{\PPL}{PPL}
\DeclareMathOperator{\Rena}{R_{1}}
\DeclareMathOperator{\Rpet}{R_{5}}

\DeclareMathOperator{\Rk}{R_{k}}
\DeclareMathOperator{\MRR}{MRR}
\DeclareMathOperator{\MRRpet}{\MRR_5}

\DeclareMathOperator{\rank}{rank}
\DeclareMathOperator{\lev}{lev}
\DeclareMathOperator{\exactmatch}{EM}
\DeclareMathOperator{\editsim}{ES}
\DeclareMathOperator{\prefixsim}{PS}

\newcommand{\R}{\mathbb R}
\newcommand{\indicator}{\mathbbm 1}
\newcommand{\vocab}{\mathbb V}
\newcommand{\retr}{\textsc{Ret}_{\mathcal{D}}}
\newcommand{\retrolayer}{\textsc{Retro}}
\newcommand{\codetp}{CodeT5+}

\newcommand{\gptsc}{\textsc{gpt}}
\newcommand{\starcodersc}{\textsc{SC}}

\newcommand{\gptSsc}{GPT$_6^\starcodersc$}
\newcommand{\gptSgpt}{GPT$_6^\gptsc$}
\newcommand{\gptMsc}{GPT$_9^\starcodersc$}
\newcommand{\gptMgpt}{GPT$_9^\gptsc$}

\newcommand{\gptLgpt}{GPT$_{12}^\gptsc$}

\newcommand{\retroSsc}{RETRO$_6^\starcodersc$}
\newcommand{\retroSgpt}{RETRO$_6^\gptsc$}

\newcommand{\retroLgpt}{RETRO$_{12}^\gptsc$}

\newcommand{\tinystarcoder}{StarCoder$_{164M}$}
\newcommand{\bigstarcoder}{StarCoder$_{15.5B}$}
\newcommand{\davinci}{code-davinci-002}

\newcommand{\n}{\string\n}
\newcommand{\ws}{\textvisiblespace}

\newcommand{\prob}{P}

\newcommand{\repoevalline}{\texttt{line}}
\newcommand{\repoevalapi}{\texttt{api}}
\newcommand{\repoevalfunc}{\texttt{function}}
\newcommand{\repoevallinerand}{\texttt{lineR}}
\newcommand{\repoevalapirand}{\texttt{apiR}}

\newcommand{\lineclean}{\texttt{line'}}
\newcommand{\apiclean}{\texttt{api'}}
\newcommand{\linerandclean}{\texttt{lineR'}}
\newcommand{\apirandclean}{\texttt{apiR'}}

\newcommand{\CI}[3]{#1$_{#2}^{#3}$}
\newcommand{\cmp}[2]{\textcolor{OliveGreen}{#1} / \textcolor{BrickRed}{#2}}

\newcommand{\offset}[1]{#1^{\geq 64}}

\newcommand{\bolje}[1]{\scriptsize{#1\%}}

\journal{Expert Systems with Applications}

\begin{document}

\begin{frontmatter}



\title{Retrieval-augmented code completion for local projects using large language models\tnoteref{t0,t1}} %% Article title


\author[1,2]{Marko Hostnik} %% Author name
% \cortext[cor1]{Corresponding author}

\affiliation[1]{organization={Faculty of Computer and Information Science, University of Ljubljana},%Department and Organization
            % addressline={Večna pot 113}, 
            % city={Ljubljana},
            % postcode={1000}, 
            % state={Ljubljana},
            country={Slovenia}}

\affiliation[2]{organization={Faculty of Mathematics and Physics, University of Ljubljana},%Department and Organization
            % addressline={Jadranska ulica 19}, 
            % city={Ljubljana},
            % postcode={1000}, 
            % state={Ljubljana},
            country={Slovenia}}

\author[1]{Marko Robnik-Šikonja} %% Author name

\tnotetext[t0]{This is the author’s accepted manuscript. The final published version is available at \url{https://doi.org/10.1016/j.eswa.2025.128596}.}
\tnotetext[t1]{© 2025. This manuscript version is made available under the CC-BY-NC-ND 4.0 license \url{https://creativecommons.org/licenses/by-nc-nd/4.0/}.}

%% Abstract
\begin{abstract}
The use of large language models (LLMs) is becoming increasingly widespread among software developers.
However, privacy and computational requirements are problematic with commercial solutions and the use of LLMs.
In this work, we focus on using relatively small and efficient LLMs with 160M parameters that are suitable for local execution and augmentation with retrieval from local projects.
% In this work, we focus on using LLMs with around 160 million parameters that are suitable for local execution and augmentation with retrieval from local projects.
%We train two models based on the transformer architecture, the generative model GPT-2 and the retrieval-adapted RETRO model, on open-source Python files, and empirically evaluate and compare them, confirming the benefits of vector embedding based retrieval.
We train two open transformer-based models, the generative GPT-2 and the retrieval-adapted RETRO, on open-source Python files, and empirically compare them, confirming the benefits of embedding-based retrieval.
Furthermore, we improve our models' performance with In-context retrieval-augmented generation (RAG), which retrieves code snippets using the Jaccard similarity of tokens.
We evaluate In-context RAG on larger models and determine that, despite its simplicity, the approach is more suitable than using the RETRO architecture.
Experimental results indicate that In-context RAG improves the code completion baseline by over 26\%, while RETRO improves over the similarly sized GPT-2 baseline by 12\%. We highlight the key role of proper tokenization in achieving the full potential of LLMs in code completion.

\end{abstract}


\end{frontmatter}



\section{Introduction}\label{sec:introduction}
% The introduction should clearly state the objectives of your work. We recommend that you provide an adequate background to your work but avoid writing a detailed literature overview or summary of your results. 

% Background
Programming with the help of large language models (LLMs) and other artificial intelligence (AI) tools is becoming increasingly widespread~\citep{copilot-survey-10.1145/3597503.3608128}. 
Programmers use AI tools (e.g., GitHub Copilot\footnote{\url{https://github.com/features/copilot}}) to suggest next lines while writing code~\citep{li2023starcoder, guo2024deepseekcoder, codellama-rozière2024code}, to write tests and shorter or longer sections of code~\citep{mbpp-austin2021program, lu2021codexglue}, and to facilitate learning new technologies~\citep{chatgpt-education-KASNECI2023102274}. 

% Problem statement
As AI tools prove increasingly useful, code editor developers are rapidly incorporating AI into their products.
However, the widespread adoption of AI tools raises concerns about code privacy, as code is transmitted to remote servers for processing by commercial models that can contain hundreds of billions of parameters~\citep{gpt3-NEURIPS2020_1457c0d6}. 
These large models require computational resources far beyond what ordinary consumer hardware provides, which is why a number of solutions that allow LLMs to run efficiently on consumer hardware are being developed.\footnote{An example is available at \url{https://github.com/ggerganov/llama.cpp}.}

% Research gap
However, the choice of a model size represents an important trade-off, as the model size has a significant impact on the quality of suggestions, the inference speed, and the utilization of compute and memory resources.
While existing approaches have focused on improving suggestions using larger models augmented with retrieval from large retrieval datasets~\citep{retro-borgeaud22}, our approach differs in using small models augmented with only locally available project files.

By focusing on a single programming language or solving just one specific programming task, such as line completion, smaller models can achieve useful results. 
Additionally, smaller models can speed up the performance of larger models through speculative decoding, where a larger model is used to correct the errors of a smaller model~\citep{speculative-decoding-pmlr-v202-leviathan23a}.
In our work, we focus on the use of LLMs with fewer parameters to complete lines of Python code. 
This allows the model to be executed within the code editor on the user's hardware, even without a graphics processing unit (GPU) to speed up the execution.

% Proposed solution
LLMs take a string as input and predict its most probable continuation.
Usually, the input string or \emph{context} is the content of the current file being edited. 
Programming projects are often composed of many files that refer to each other, so we want to enrich the context with information from other files in the project.
However, as the context size of models is limited from a few dozen to a hundred lines of code, we need to find informative data to include in the context.
In our work, we augment the context of LLMs with relevant code snippets from other files in the project, which we search for based on the similarity of vector embeddings, or alternatively based on the Jaccard similarity of tokens.
The goal of our work is to experimentally verify the impact of retrieval-augmented LLMs on the success of completing lines of code. We focus on smaller models suitable for inference without a GPU and on retrieval only from local projects. 

% Highlight how the proposed approach can improve the coding experience for developers
Our proposed approach has widespread use cases in improving the coding experience for developers. 
Small coding LLMs can be used to offer intelligent code completion suggestions in various environments, such as in local text editors.
In our work, we demonstrate how small LLMs can be improved and made more viable with retrieval-augmented generation (RAG) using a simple Jaccard similarity-based approach that doesn't require expensive indexing for retrieving relevant code snippets from developers' projects.
Better code suggestions with smaller LLMs can not only help developers write code faster, but our approach does so while maintaining a focus on privacy without incurring additional costs of commercial LLMs.

% Contributions
Our main contributions are as follows:
\begin{itemize}
    \item We train and compare 160 million parameter RETRO~\citep{retro-borgeaud22} and GPT-2~\citep{gpt2-radford2019language} models using two different tokenizers and verify the importance of token healing for code completion.
    % \item We confirm that the 160M parameter RETRO model with retrieval from local projects outperforms the similarly sized baseline model.
    \item \sloppy We compare In-context RAG with Jaccard-similarity-based retrieval to RETRO with embedding-based retrieval and to baseline models of different sizes.
    \item We examine the impact of copying from code snippets and the impact of the retrieved snippets' similarities to the retrieval query.
    \item We experimentally confirm the viability of augmenting small LLMs with retrieval from local projects for local use in code editors.
\end{itemize}
This paper is organized into six sections.
In Section~\ref{sec:related-work}, we present related work in retrieval-augmented code completion.
In Section~\ref{sec:methods}, we present the main methods used, including an overview of the RETRO architecture~\citep{retro-borgeaud22}.
In Section~\ref{sec:experimental-setup}, we describe the setup of our experiments, including the dataset, model training details, and metrics used.
In Section~\ref{sec:results}, we present and analyze the results of our trained models evaluated with retrieval from local projects.
We conclude in Section~\ref{sec:conclusion}, where we also present ideas for further work.

\section{Related work}\label{sec:related-work}

The field of LLMs based on the transformer architecture~\citep{attention-is-all-you-need} has been rapidly developing in recent years, not only for natural language processing but also for the processing of programming code. We present an overview of related work, split into four subsections: LLMs for code, retrieval-augmented LLMs, retrieval-augmented code completion, and project based evaluation of code completion.

\subsection{Large language models for code}
Code generation is important for tasks such as code completion~\citep{li2023starcoder, guo2024deepseekcoder, codellama-rozière2024code}, 
generating code from natural language instructions~\citep{mbpp-austin2021program}, 
% code summarization~\citep{gnn-code-summarization-liu2021retrievalaugmented}, 
translating code from one programming language to another~\citep{lu2021codexglue}, 
and code searching based on natural language or code-based queries~\citep {wang-etal-2023-codet5}. 
Neural models for code completion are mostly based on the decoder part of the transformer architecture and differ mainly in the number of parameters, the size of the context and the training data used (e.g., the proportion of each programming language and inclusion of natural language)~\citep{li2023starcoder, guo2024deepseekcoder}. 
When training LLMs with longer context sizes, concatenating project files  into the input context is becoming increasingly common~\citep{guo2024deepseekcoder, codellama-rozière2024code}. 
This better prepares the model for including related files into the context during model execution. 
% Some approaches augment the input data with the structure of the code~\citep{CodeFill-10.1145/3510003.3510172} and use graph neural networks for processing~\citep{gnn-code-summarization-liu2021retrievalaugmented, gnn-allamanis2018learning}.

\subsection{Retrieval-augmented language models}
The use of retrieval in LLMs is often motivated by the desire to improve models' capabilities in open-domain question answering tasks~\citep{nlp-transformers-info14040242}.
DPR~\citep{dpr-karpukhin-etal-2020-dense} and REALM~\citep{realm-pmlr-v119-guu20a} extract the answer from the retrieved data using a BERT encoder~\citep{bert-devlin2019bert}. 
DPR fine-tunes the weights of the query encoder to better match its outputs with the frozen document encoder embeddings. 
REALM updates the weights of the knowledge retriever encoder during training, which causes computationally demanding re-indexing of data during training.
RAG~\citep{RAG-NEURIPS2020_6b493230} and FiD~\citep{FiD-izacard:hal-03463108} differ in the processing of the retrieved snippets with additionally fine-tuned encoder-decoder models. 
\citet{incontext-ralm-10.1162/tacl_a_00605} show that retrieval is beneficial even without additional training of models; it is sufficient to append the found snippets to the context of the decoder. 
$k$NN-LM~\citep{knn-Khandelwal2019GeneralizationTM} computes a weighted sum of the predictions from an LLM and a $k$-nearest neighbors classifier when predicting each token. 
% GNN-LM~\citep{gnn-lm-meng2021gnn} enhances $k$NN-LM by using graph neural networks to combine the retrieved snippets.
RETRO~\citep{retro-borgeaud22}, described in Section~\ref{sec:retro}, performs chunk-based retrieval and includes the found chunks in the decoder via cross-attention, which, in combination with a frozen document encoder, enables retrieval throughout the entire training of the model.

\subsection{Retrieval-augmented code completion}
Approaches for retrieval-augmented code completion are similar to those for natural language based open-domain question answering, with differences in the use of retrieval adapted to code and projects as a whole. 
ReACC~\citep{ReACC-lu2022reacc} (and similarly REDCODER~\citep{redcoder-parvez-etal-2021-retrieval-augmented}) combines retrieval with a code-adapted encoder, GraphCodeBERT~\citep{graphcodebert}, and classic BM-25 search~\citep{bm25-harman1995overview}, incorporating found snippets into the context. 
% $k$NM-LM~\citep{knm-lm-10298575} adapts $k$NN-LM for the domain of code using Bayesian statistics. 
\citet{pmlr-v202-shrivastava23a} use a trained classifier to select the most relevant code snippet from the project to include into the context. 
RepoCoder~\citep{repocoder-zhang-etal-2023-repocoder} demonstrates retrieval based on the Jaccard similarity of tokens for retrieving line-based snippets from the project and utilizes the generated code to improve subsequent retrieval. 
This approach is the motivation for our approach described in Section~\ref{sec:incontext-jaccard}. 
RepoFormer~\citep{wu2024repoformer} enhances RepoCoder with adaptive retrieval triggered by the model with a special token.
DRAG~\citep{shapkin2023entity} extends the model's vocabulary with new tokens from entities in the project that the model can use while generating code.

\subsection{Project based evaluation of code completion}
There are several datasets for evaluating code completion. 
HumanEval \citep{humaneval-chen2021evaluating} involves completing functions from given function names and docstrings; CodeXGLUE~\citep{lu2021codexglue} contains many datasets for various tasks, including line completion. 
These datasets only evaluate general code completion abilities within a single file, whereas real projects contain multiple interrelated files. 
Consequently, datasets for project-level completion have begun to emerge, often as an addition for evaluating the developed approach~\citep{shapkin2023entity}. 
RepoBench~\citep{liu2023repobench} and CrossCodeEval~\citep{crosscodeeval-NEURIPS2023_920f2dce} compile evaluation sets by pre-selecting useful code snippets from related files that the model should include in the context. 
On the other hand, RepoEval~\citep{repocoder-zhang-etal-2023-repocoder} -- more in Section~\ref{sec:repoeval-dataset} -- contains entire projects in addition to the marked lines that need to be completed, which allows the use of custom implementations of retrieval from the project and consequently any model. % (e.g., RETRO or $k$NM-LM would be difficult to use with a dataset like CrossCodeEval). 

\section{Code line completion methods}\label{sec:methods}
% The materials and methods section should provide sufficient details about your materials and methods to allow your work to be reproduced by an independent researcher.
In this section, we describe our methodology for code line completion based on LLMs. 
We first describe the retrieval-enhanced RETRO approach, followed by our improvements, the first concerning the tokenization, and the second using context-based retrieval-augmented generation.

\subsection{The RETRO approach}\label{sec:retro}
\begin{figure*}[h]
    \centering
    \includegraphics[width=0.66\linewidth]{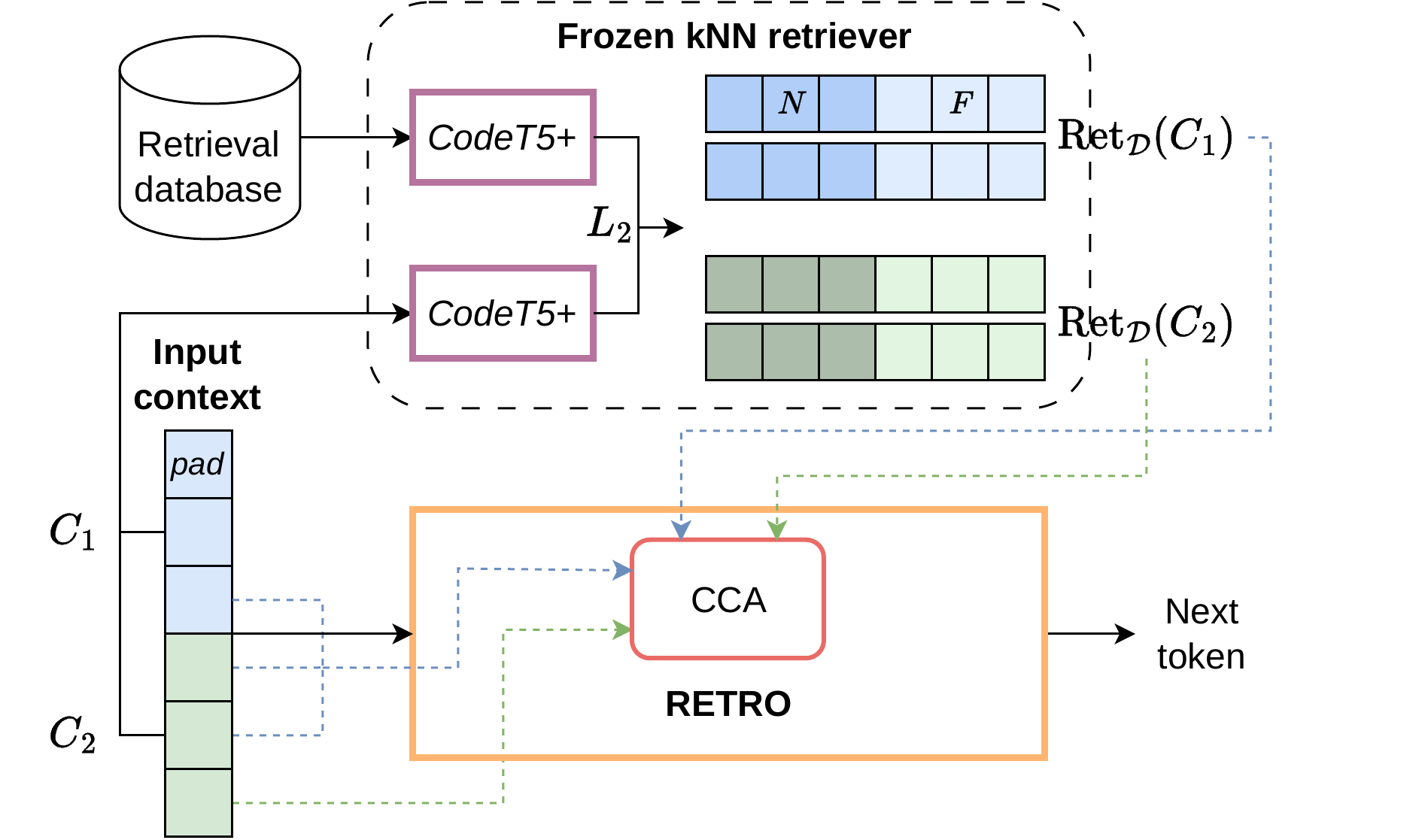}
    \caption{
    Illustration of our RETRO approach.
    The CodeT5+~\citep{wang-etal-2023-codet5} encoder is used for computing chunk embeddings for retrieval based on the $L_2$ distance.
    The color coded arrows indicate which tokens attend to which retrieved chunks -- refer to \citet{retro-borgeaud22} for more details on the chunked cross-attention (CCA) mechanism.
    During sampling, the input is padded to be a multiple of the chunk size.
    }
    \label{fig:retro}
\end{figure*}

RETRO (Retrieval-Enhanced Transformer) is a transformer based autoregressive language model enhanced with additional snippets retrieved from a database~\citep{retro-borgeaud22}. 
The approach splits the input token sequence into contiguous fixed-size chunks and finds snippets similar to the previous chunk when processing the current chunk.
Introducing retrieval into the model allows adapting the model to new data without updating its weights, which is particularly useful for predicting facts not present in the training set (e.g., names in a local programming project). 
For our experiments, we make minor modifications to the original RETRO architecture.
We change the chunk embedding model from BERT~\citep{bert-devlin2019bert} to CodeT5+~\citep{wang-etal-2023-codet5} and modify the hyperparameters described in Section~\ref{sec:model-architecture}.
During inference, we modify the input context by adding some padding tokens, using token healing (see Section~\ref{sec:token-healing}) and In-context RAG (see Section~\ref{sec:incontext-jaccard}).
Our RETRO approach is illustrated in Figure~\ref{fig:retro}.
Below, we explain the details of the approach, relevant for our application.

\paragraph{Language modeling with retrieval}

We denote the token vocabulary by the set $\vocab$ obtained by a text tokenizer (see Section~\ref{sec:tokenization}).
The input token sequence $x = (x_1, ..., x_n) \in \vocab^n$ of length $n \in \mathbb{N}$ is split into $l \in \mathbb{N}$ contiguous chunks $C_1 = (x_1, ..., x_m), ..., C_l = (x_{n-m+1}, ..., x_n)$ of length $m = 64$. 
The probability of the next token $x_i$ depends only on previous tokens and snippets retrieved by previous chunks:
\begin{equation*}
    \prob \left( x_i \mid \theta, (x_j)_{j<i} , \left( \retr(C_v) \right)_{v < u_i} \right),
\end{equation*}
where $u_i = \lceil \frac{i}{m} \rceil$ is the index of the chunk containing $x_i$, $\left( \retr(C_v) \right)_{v < 1} = \emptyset$, $\theta$ are the weights of the model, and
$\retr(C_u)$ denotes the $k$ retrieved snippets for $C_u$ from the retrieval database $\mathcal{D}$. 
% The retriever $\retr$ has no parameters updated during training. 

% A peculiarity of the RETRO model architecture is that when predicting token $x_i$ in chunk $C_{u_i}$, the model has direct access only to the retrieved snippets of the previous chunk $\retr(C_{u_i - 1})$ in addition to $x_1, ..., x_i$. 
% However, since token $x_j \in C_{u_i - 1}$ has access to $\retr(C_{u_j - 1})$, $x_i$ has indirect access to the snippets of all previous chunks $C_1, ..., C_{u_i - 1}$ due to self-attention.

\paragraph{Retrieval}\label{sec:knn-retrieval}

% The retrieval database is built with RETRO's chunk-based architecture in mind, which splits the input into fixed size chunks.
The retrieval database is built by splitting the set of input documents $\mathcal{D}$ into fixed-size chunks of $m = 64$ tokens. 
The database stores key-value pairs, where the key is the embedding of a chunk of tokens $N$, and the value is a pair $[N, F]$, which contains both the chunk $N$ and the next immediate chunk $F$ in the original document.
The retriever $\retr(C)$ finds the $k$ most similar snippets by comparing the $L_2$ distance between the embedding of the chunk $\mathcal{M}(C)$ and the database key $\mathcal{M}(N)$.
\citet{retro-borgeaud22} use the BERT model~\citep{bert-devlin2019bert} for $\mathcal{M}$, while we use the CodeT5+ encoder~\citep{wang-etal-2023-codet5}, as explained in Section~\ref{sec:retrieval-database}.
RETRO thus receives as input the input tokens $x$ and the $k$ nearest snippets of each input chunk $\retr(C_u) = ([N^1, F^1], ..., [N^k, F^k])$. 
% , which is a total of $k \cdot 2m$ additional tokens for each chunk $C_u$. 
When training the model, we ensure that $\retr(C_u)$ does not contain $C_{u+1}$ by filtering out chunks $[N, F]$ that originate from the same training document as $x$.

\paragraph{The RETRO architecture}\label{sec:retro-architecture}

The RETRO architecture is based on the encoder-decoder transformer architecture~\citep{attention-is-all-you-need}. 
It differs from the decoder-only GPT-2 architecture by the addition of a bidirectional encoder and a chunked cross-attention mechanism, which operates on chunks and allows the inclusion of retrieved snippets into the decoder. 
Splitting the input context into fixed-length chunks enables efficient chunk-based retrieval, providing more relevant examples for the entire context and allowing for the encoder to independently process all $k$ retrieved snippets of chunks. 
Since the retriever itself is not updated during training, it can be updated after the training is complete without modifying the model weights. 
The RETRO model introduces $\retrolayer$ layers, which replace certain regular decoder layers. 
Following \citet{retro-borgeaud22}, we place $\retrolayer$ layers at the sixth layer and then at every subsequent third layer (e.g., at layers $6, 9, 12$, if there are 12 layers in total). 

\paragraph{Sampling with RETRO}\label{sec:retro-generation}

Just as the architecture of RETRO influences the construction of the retrieval database and chunked cross-attention, it also impacts sampling from the model.
When sampling from RETRO, we pad the context to make sure it is a multiple of the chunk size $m$~\citep{retro++-wang2023shall}.
This allows RETRO to retrieve based on the last $m$ tokens in the context, as otherwise predicting within the first chunk would not be improved by retrieval and tokens in the last chunk of the context would not be used for retrieval despite being most relevant to predicting the next token.
When generating longer texts, an additional question is how often to perform retrieval.
In our experiments, we limit the task to code completion until the end of a line, so retrieval is performed only before the first step of text generation or whenever the context length reaches a multiple of the chunk size.
% It should be noted that retrieval could also be performed at each step, but this would require frequent padding of the context.

% Tokens in the first chunk of the input context are used to search for a similar chunk from the retrieval database, which is then used for predicting next tokens in the second chunk. 
% However, no retrieved data is used for making predictions within the first chunk.
% Consequently, if the input sequence is shorter than the chunk size $m$, retrieval will not improve predictions. 
% This issue is solved by padding the context with special tokens, as shown in Figure~\ref{fig:retro-padding}.

% \begin{figure}[htb]
% \centering
% \includegraphics[width=0.9\linewidth]{retro++_padding.png}
% \caption{Left padding of the context when sampling from the RETRO model. Adapted from~\citet{retro++-wang2023shall}.}
% \label{fig:retro-padding}
% \end{figure}

% A similar issue occurs when we want to sample text with an input context that is not a multiple of the chunk size. 
% In this case, retrieval does not use information from tokens closer to the end of the context, even though we expect these tokens to contain the most important information for retrieving related chunks. 
% % In the extreme case, when the context length equals $c \cdot m - 1$ for $c\in \mathbb{N}$, the last $m - 1$ tokens are not used for retrieval at all. 
% The solution to this problem is to either trim or pad the context from the left, as shown on the right side of Figure~\ref{fig:retro-padding}.

\subsection{Token healing}\label{sec:token-healing}

\begin{figure*}[tb]
    \centering
     \begin{subfigure}[b]{0.49\textwidth}
         \centering
         \includegraphics[width=\textwidth]{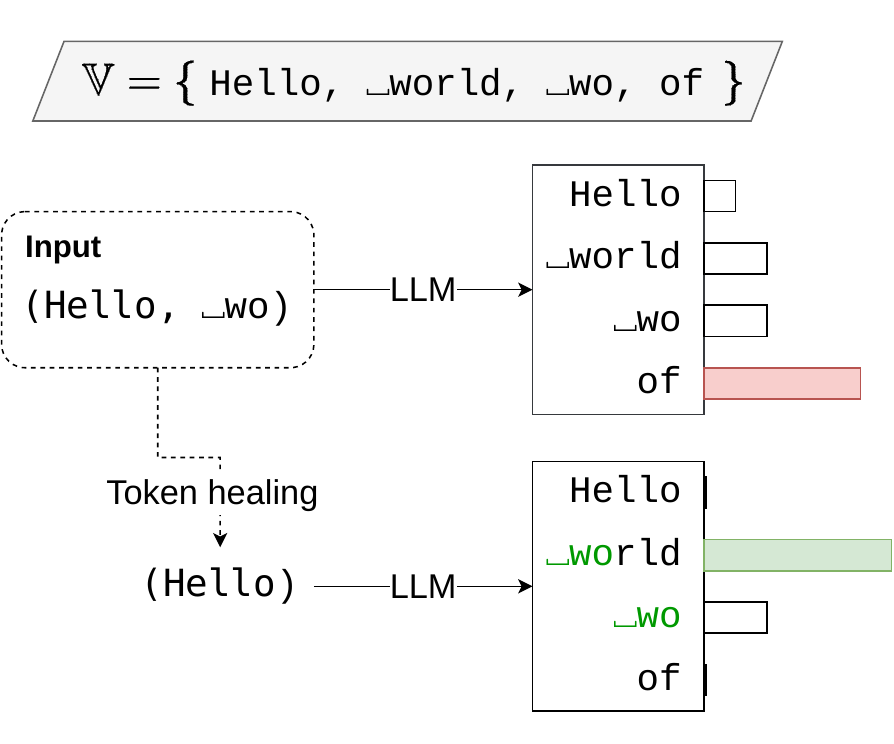}
         \caption{Token healing example.}
         \label{fig:token-healing-diagram}
     \end{subfigure}
     \begin{subfigure}[b]{0.49\textwidth}
         \centering
         \includegraphics[width=\textwidth]{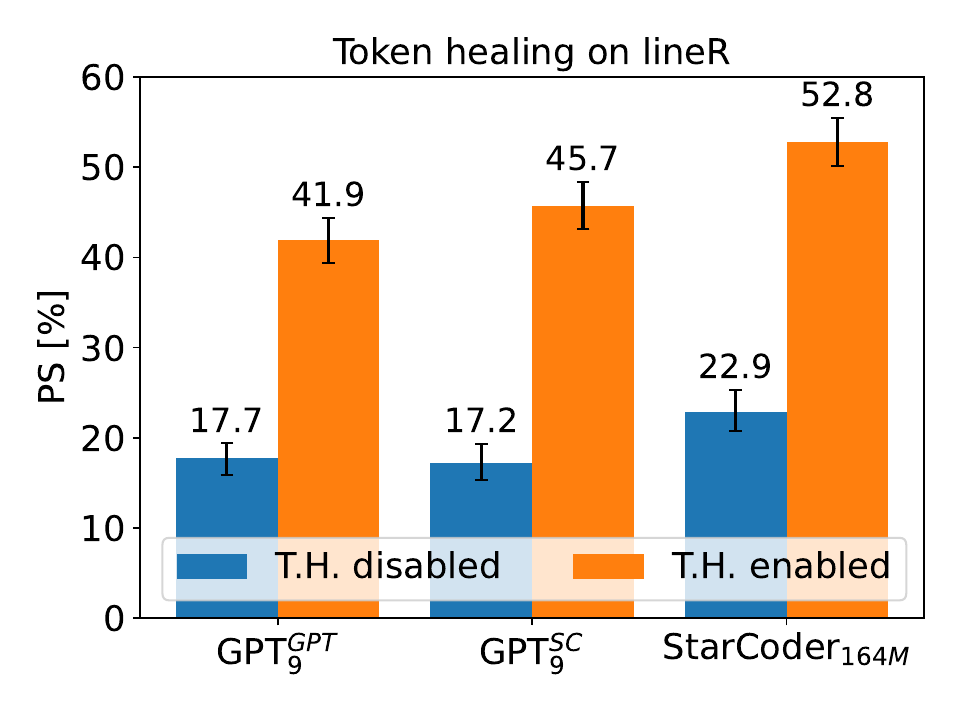}
         \caption{Metrics with or without token healing (T.H.).}
         \label{fig:token-healing-metrics}
     \end{subfigure}
    \caption{
        Figure~\ref{fig:token-healing-diagram} shows an example of token healing using a small four token vocabulary.
        With token healing, the input is rolled back by one token and the output distribution is constrained to tokens that start with \texttt{{\ws}wo}, improving over the prediction without token healing.
        % Without token healing, the LLM's output distribution
        In Figure~\ref{fig:token-healing-metrics}, we compare code completion results (without retrieval) with respect to the use of token healing using three different 160 million parameter LLMs that we evaluate in Section~\ref{sec:results}.
        The experiments are performed on the evaluation set \repoevallinerand\ described in Section~\ref{sec:repoeval-dataset}. 
        The Prefix Similarity ($\prefixsim$) metric is defined in Section~\ref{sec:metrics} (higher values are better).
    }
    \label{fig:token-healing}
\end{figure*}

The use of subword tokenization can, in certain cases, lead to undesirable effects during text generation, especially when completing lines of code starting at unnatural word boundaries.
Due to the commonly used byte pair encoding (BPE) tokenization algorithm~\citep{bpe-sennrich-etal-2016-neural}, some substrings appear in multiple tokens.
Consequently, if the input string $s$ ends with a substring that is the beginning of some token, the output probability distribution for the next token will be biased~\citep{token-healing-dagan2024getting}. 
The tokenization of the string $s$ may end with a token that would be merged with the token we are predicting in the BPE algorithm.
As a result, tokens that continue the tokenized string $s$ have a higher probability than tokens that would better complete the string $s$ truncated by the last token.
In addition, without a corrective measure, called token healing, $s$ may be an out-of-distribution example, leading to unpredictable completion results, since the model has not encountered such unnatural cut-off boundaries during training.

We solve this problem by finding and removing the longest suffix $t$ in the input string $s = s' \cdot t$ that is a prefix of some token $x_i \in \vocab$, $x_i = t \cdot x_i'$~\citep{token-healing-dagan2024getting, athiwaratkun2024token}.
During decoding, we then constrain the output probability distribution for the next token only to tokens that match the prefix of $t$.
When we generate $t$ in its entirety, we no longer constrain the output distribution.
We illustrate the token healing process in Figure~\ref{fig:token-healing-diagram} and show its practical impact in Figure~\ref{fig:token-healing-metrics}.
We observe a significant impact largely due to the choice of the evaluation dataset, which begins code completion at a random position within the line.

% \begin{table}[h]
%   \centering
%   \padtable
%   \begin{tabular}{cc|ccc}
%       Model & Token healing & $\exactmatch$ & $\editsim$ & $\prefixsim$  \\
%       \hline
%       \multirow{2}{*}{\gptMgpt} & No &\CI{17.8}{16.1}{19.8} & \CI{46.5}{45.0}{48.3} & \CI{17.7}{15.9}{19.4} \\ 
%       & Yes & \CI{41.9}{39.6}{44.2} & \CI{66.1}{64.4}{67.6} & \CI{41.9}{39.4}{44.4}  \\  
      
%       \hline
%       \multirow{2}{*}{\gptMsc} & No & \CI{16.8}{15.2}{18.7} & \CI{46.8}{45.3}{48.6} & \CI{17.2}{15.3}{19.3} \\ 
%       & Yes & \CI{45.2}{42.8}{47.5} & \CI{68.3}{66.6}{69.9} & \CI{45.7}{43.1}{48.4}  \\ 
      
%       \hline
%       \multirow{2}{*}{\tinystarcoder} & No & \CI{23.4}{21.7}{25.8} & \CI{53.1}{51.5}{55.0} & \CI{22.9}{20.7}{25.3} \\
%       & Yes & \CI{52.2}{49.7}{54.6} & \CI{72.6}{71.1}{74.2} & \CI{52.8}{50.1}{55.5} \\
%   \end{tabular}
%   \caption{
%   Comparison of code completion results (without retrieval) with respect to the use of token healing.
%   The experiments are performed on the evaluation set \repoevallinerand\ described in Section~\ref{sec:repoeval}. 
%   The metrics $\exactmatch$, $\editsim$ in $\prefixsim$ are defined in Section~\ref{sec:metrics} (higher values are better).
%   }
%   \label{tab:token-healing}
% \end{table}

\subsection{In-context RAG}\label{sec:incontext-jaccard}

A simpler alternative for retrieval-augmented code completion, compared to RETRO, is to include similar code snippets directly into the model's context. 
This eliminates the additional complexity of the RETRO architecture, as the In-context RAG method can be used with any autoregressive LLM~\citep{incontext-ralm-10.1162/tacl_a_00605, repocoder-zhang-etal-2023-repocoder, ReACC-lu2022reacc}.

\begin{figure*}[htb]
    \centering
    \includegraphics[width=0.8\textwidth]{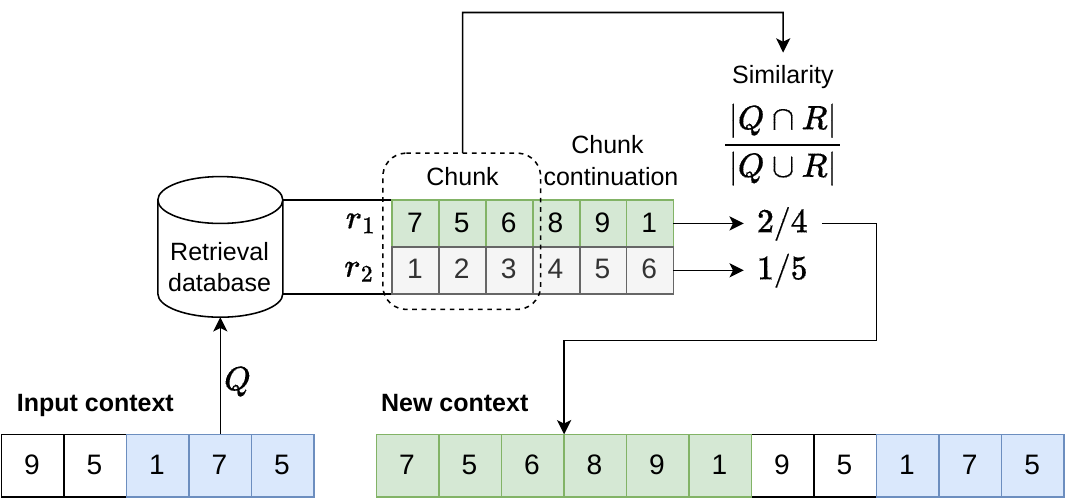}
    \caption{Our In-context RAG approach based on the Jaccard similarity of chunks of tokens.
    The retrieved snippet contains continuation tokens and additional metadata about the origin file, while the key for comparing with the query is a fixed length chunk.
    }
    \label{fig:rag-jaccard}
\end{figure*}

Let $x=(x_1, ..., x_n) \in \vocab^n$ be the input sequence of tokens.
From $x$, we construct a query $q$ from the last $m \in \mathbb{N}$ tokens, $q = (x_{n-m+1}, ..., x_n)$, and use it to search for the $k$ nearest code snippets from the database $\mathcal{D}$.
The tokens of the found snippets $\retr(q) = (r_1, ..., r_k)$ are concatenated into a sequence of tokens $r = r_1 \cdot ... \cdot r_k$. 
The new context for the model is then $x' = r \cdot x$, which is used to predict the next tokens.
The process is illustrated in Figure~\ref{fig:rag-jaccard}.

Individual retrieved snippets $r_i$ can be further enhanced, for example by adding metadata about the original file, such as $r_i' = \texttt{\# path/to/file.py\n} \cdot r_i$. 
Unlike RETRO, this approach is not limited to retrieving fixed size chunks.
We can also dynamically adjust the number of retrieved snippets $k$ based on the remaining size of the input context.
The retrieval database $\mathcal{D}$ and the query $q$ need not be constructed based on fixed-length chunks of tokens; instead, the data can be chunked based on lines or logical code parts, such as functions.
In our experiments, to make a fair comparison, we use fixed chunks as in the RETRO approach.
As with RETRO, the retriever $\retr$ only returns examples that are not from the same file as $x$.

% \paragraph{Retrieving based on the Jaccard similarity coefficient}
As with the RETRO approach, the retriever $\retr(q)$ could operate based on the distance between the embedding of the query and chunks from the database $r_i \in \mathcal{D}$, $d(\mathcal{M}(r_i), \mathcal{M}(q))$.
We decided on a simpler approach that works solely based on the Jaccard similarity of tokens and not on embeddings. 
The Jaccard similarity coefficient $J$ is used over the set $Q$ of tokens from the query $q$ and the set $R$ of tokens from the retrieved snippet $r_i$,
\begin{equation*}
J(Q, R) = \frac{|Q \cap R|}{|Q \cup R|}.
\end{equation*}
The retriever $\retr$ thus finds code snippets that have the highest similarity between the query $q$ and the chunks in the database $\mathcal{D}$.
We prefer this approach because it does not require an additional model to compute embeddings, making it more suitable for efficient execution on limited hardware and with local projects.

\section{Experimental settings }\label{sec:experimental-setup}
In this section, we present the datasets, evaluation metrics, and experimental settings used to verify if retrieval-augmented code completion can enhance the performance of LLMs when working on local projects.
Since we aim to run models on local hardware, we focus on smaller models and efficient retrieval and text generation methods. 
We assess the impact of retrieval by comparing GPT-2, RETRO, and the In-context RAG approach that incorporates found snippets into the context of models without modifying the model architecture or training process. We split the section into seven parts: tokenization, training and evaluation dataset descriptions, retrieval database, model architecture with hyperparameters, training of models, and used comparison metrics. The results are reported in Section \ref{sec:results}.

\subsection{Tokenization}\label{sec:tokenization}

We compare two subword tokenization methods based on the BPE algorithm~\citep{bpe-sennrich-etal-2016-neural} with differing vocabularies. 
The first version of the vocabulary with \num{50257} tokens was trained on a diverse textual dataset obtained from web sources used in the GPT-2 paper~\citep{gpt2-radford2019language}. 
The second version of the vocabulary with \num{49152} tokens was trained on a collection of open-source code from various programming languages reported in the StarCoder paper~\citep{li2023starcoder}.
Since we work on the code completion task, the use of tokenization from the StarCoder paper shall be more appropriate than using a vocabulary more suitable for natural language processing.

\begin{table}[htb]
    \centering
    \begin{tabular}{@{}ll@{}}
         \toprule
         Tokenization & Tokens  \\
         \midrule
         GPT-2 & $(\texttt{def},\, \texttt{\ws foo},\, \texttt{():},\, \texttt{\n},\, \texttt{\ws},\, \texttt{\ws},\, \texttt{\ws},\, \texttt{\ws pass})$ \\
         StarCoder & $(\texttt{def}$, $\texttt{\ws foo}$, $\texttt{():}$, $\texttt{\n\ws\ws\ws}$, $\texttt{\ws pass})$ \\
         \bottomrule
    \end{tabular}
    \caption{Tokenizations of the string $\texttt{def foo():\n\ws\ws\ws{\ws}pass}$. Tokens are comma separated. The StarCoder tokenization requires fewer tokens due to the merging of whitespace characters into a single token.}
    \label{tab:tokenization-example}
\end{table}

The main drawback of using the text-based GPT-2 vocabulary for code completion is an inefficient handling of whitespaces. 
In most cases, each space in a longer sequence of spaces is represented by its own token, which significantly reduces the number of useful tokens available in the model's context during the self-attention computation.
% This is problematic because the Python programming language typically uses four spaces (less frequently tabs \verb|\t|) for each level of indentation.
An example is in Table~\ref{tab:tokenization-example}.
In addition to less useful data in the model's context, the inefficient tokenization of consecutive spaces also affects the amount of information in each chunk of the retrieval database, as the database is built with a fixed number of tokens per chunk.

\subsection{The StarCoder dataset}\label{sec:dataset}

For training our models, we use the StarCoder dataset\footnote{
    Available at \url{https://huggingface.co/datasets/bigcode/starcoderdata}. 
    % At the time of writing, StarCoder2 has also been released, which we do not use.
}~\citep{li2023starcoder}. 
The dataset contains over 300 million open-source files from more than 80 programming languages, but we use only files in the Python programming language.
% Each example in the dataset contains the name of the project from which the file originates, the full path of the file within the project, and the file content itself.
We split the dataset into a training and test set in ratio 90\%; 10\%.
We split the data at the project level and not at the file level, so that all files belonging to a project are placed in the same training or test set. 
The validation set is constructed from 1\% of the test set and is used to monitor metrics during model training to detect any potential overfitting (which we did not observe).
% We do not use the validation set to optimise the hyperparameters as we do not have the computational capacity for such experiments.
After training, we evaluate the models' performance on both the test set and the evaluation sets as described in Section~\ref{sec:repoeval-dataset}. 
Table~\ref{tab:starcoderdata} summarizes the sizes of the training and test sets.

\begin{table}[htbp]
\centering
\begin{tabular}{@{}lrrr@{}}
\toprule
Split & \# Files & \# Projects & Total file size \\
\midrule
Training set & \num{11579885} & \num{1510155} & 56 GiB \\
Test set & \num{1286653} & \num{168449} & 6 GiB \\
\midrule
Total & \num{12866538} & \num{1678604} & 62 GiB \\
\bottomrule
\end{tabular}
\caption{Size of the training and test sets composed from the Python part of the StarCoder dataset.}
\label{tab:starcoderdata}
\end{table}

\subsection{The RepoEval dataset}\label{sec:repoeval-dataset}
For the evaluation of retrieval-augmented line completion on real projects, we use the RepoEval dataset~\citep{repocoder-zhang-etal-2023-repocoder}, which contains eight open-source projects from various domains and is divided into three evaluation sets: \repoevalline, \repoevalapi, and \repoevalfunc.
% The \repoevalline\ and \repoevalapi\ sets each contain 1600 examples, with 200 examples from each project.
The \repoevalline\ set contains randomly selected individual lines, while the \repoevalapi\ set contains lines with application programming interface (API) calls to objects defined within the project.
These calls can be longer than a single line (10\% of examples contain at least 5 lines).
The \repoevalfunc\ set is intended for completing function bodies, but we do not use it as it requires completing up to 30 lines and evaluation is performed by executing the projects' unit tests. 

\begin{figure*}[htb]
    \centering
     \includegraphics[width=0.6\textwidth]{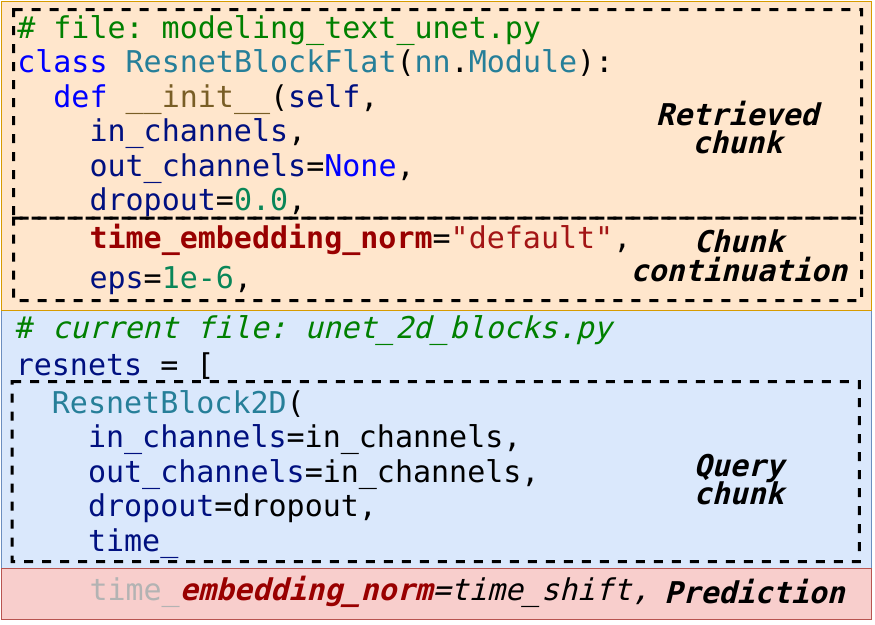}
    \caption{
    An example of the In-context RAG approach described in Figure~\ref{fig:rag-jaccard} on our \repoevallinerand\ subset of RepoEval starting completion from a random position in the target line.
    The example highlights the importance of including chunk continuation tokens along with the retrieved chunk.
    }
    \label{fig:repoeval-jaccard-example}
\end{figure*}

In both cases, \repoevalline\ and \repoevalapi, the goal is to predict the entire next line or function call.
We want to use our models in a code editor even when the cursor is in the middle of a line, while the user is typing.
Therefore, we also evaluate the models' performance on \repoevallinerand\ and \repoevalapirand, where we use the model to complete code from a random position within the line to the end of the original example.
More details on \repoevallinerand\ and \repoevalapirand\ can be found in \ref{sec:repoeval-dataset-details} and an example in Figure~\ref{fig:repoeval-jaccard-example}.

When evaluating LLMs, there is often the issue of the evaluation set being present in the training set, as training sets are frequently obtained by scraping massive amounts of data from the web.
Therefore, we check the overlap between RepoEval and our StarCoder training dataset and find that three of the eight projects in RepoEval are also present in the training set (the project with the most overlap is \textit{alibaba/FederatedScope}, with 48\% of its files contained in StarCoder).
In the resulting \lineclean, \linerandclean, \apiclean, and \apirandclean\ datasets, we remove these three projects that overlap with the training set.

% \paragraph{Experimental setup}\label{sec:repoeval-setup}
When simulating and evaluating line completion using the RepoEval dataset, the input context for models without retrieval contains only the lines before the target line within a single file.
With retrieval, the retrieval database is built only from individual projects (we do not use the entire StarCoder training set).
In our experiments, all text is generated using greedy decoding to achieve determinism and faster execution when completing short lines of code.

\subsection{Retrieval database}\label{sec:retrieval-database}
To train the RETRO approach, we need a database to search for the $k$-nearest chunks of each input chunk of tokens.
In line with \citet{retro-borgeaud22} and \citet{retro++-wang2023shall}, we use the training dataset as the retrieval database.
This ensures a fair comparison between RETRO, which uses retrieval during training, and GPT-2, which does not. Using any other additional dataset, RETRO would have an advantage, as it would have access to a larger amount of data during training.

\sloppy
For building the retrieval database, we use a chunk size of $m=64$ tokens and the CodeT5+~\citep{wang-etal-2023-codet5} encoder for computing chunk embeddings.
To enable faster retrieval of the $k$-nearest chunks, we index the stored embeddings with Faiss~\citep{faissgpu-johnson2019billion} as described in \ref{app:indexing}.
The pretrained CodeT5+ encoder has 110 million parameters, generates normalized 256-dimensional embeddings and was trained on various code tasks.
We also experimented with using the BERT-large encoder with 340 million parameters and 1024-dimensional vector embeddings, as done by~\citet{retro-borgeaud22} and described in~\ref{app:bert-codet5p}. 
However, we opt for the CodeT5+ encoder since it has fewer parameters, outputs smaller dimensional embeddings, and is trained for code understanding, making it a better fit for our use case.

% \paragraph{Chunk embeddings}\label{sec:embeddings-codetp}
% To compute the vector embeddings of chunks for constructing the retrieval database, we use the pretrained CodeT5+ encoder~\citep{wang-etal-2023-codet5}, which was trained on various code tasks. 
% The encoder has 110 million parameters and generates normalized 256-dimensional vector embeddings for the input sequence. 
% We also experimented with using the BERT encoder, which has 340 million parameters and 1024-dimensional vector embeddings, as done by~\citet{retro-borgeaud22} and described in~\ref{app:bert-codet5p}. 
% However, we opt for the CodeT5+ encoder since it has fewer parameters, produces smaller dimension embeddings, and is trained for code understanding, making it a better fit for our use case.

% \subsection{Model training}

% In this work, we limit ourselves to training smaller models that are suitable for execution on personal computers.

\subsection{Model architecture}\label{sec:model-architecture}

In accordance with \citet{retro-borgeaud22} and \citet{retro++-wang2023shall}, we compare GPT-2 with RETRO, keeping the architecture of both models identical except that RETRO includes both an encoder and decoder compared to GPT-2 which is a decoder only model.
Table~\ref{tab:model-sizes} summarizes the number of parameters in our models as well as the number of decoder layers.
To compare models with roughly equal number of parameters, we also compare GPT-2 with 9 decoder layers to RETRO with 6 decoder layers; this comparison shall ensure that any potential benefit of RETRO does not arise simply from increasing the number of parameters when adding the chunk encoder. 
% Therefore, we increase the number of layers in the GPT-2 decoder to match the number of parameters with the RETRO model as closely as possible.
\begin{table}[htb]
\centering
\begin{tabular}{@{}rrr@{}}
\toprule
\# Decoder layers & GPT-2 & RETRO \\
\midrule
6  & 126 M & \textbf{160 M} \small{(+30\%)} \\
9  & \textbf{164 M} & /      \\
12 & 201 M & 243 M \small{(+21\%)} \\
\bottomrule
\end{tabular}
\caption{
The number of parameters of our GPT-2 and RETRO models when using GPT-2 tokenization. 
The relative change in the number of parameters is shown in parentheses. 
% The relative difference between the bold values is 2.5\%. 
When using the StarCoder tokenization, the number of parameters for all models decreases by a constant two million parameters due a smaller vocabulary, which affects the number of parameters in the transformer's embedding table.
}
\label{tab:model-sizes}
\end{table}

\paragraph{Hyperparameters} We set the hidden dimension $d$ to 1024, and the number of heads in the multi-headed attention to 16, so that each head processes vectors of dimension 64. 
The number of decoder layers is shown in Table~\ref{tab:model-sizes}.
\retrolayer\ layers with the retrieval capability are placed at the sixth layer and then at every subsequent third layer (e.g., layers $6, 9, 12$ if there are 12 layers). 
The encoder in RETRO has two layers.
The maximum length of the input context is limited to 384 tokens as in~\citet{semenkin2024context}. 
Instead of fixed absolute positional embeddings, we use relative positional embeddings RoPE~\citep{rope-positional-embeddings-2021arXiv210409864S} with a rotational factor of $0.5$.
For the nonlinear function within fully connected feed-forward networks, we use the $\swiglu$ function, which has been shown to contribute to better learning outcomes in language modeling tasks~\citep{swiglu-shazeer2020glu, llama1-touvron2023llama}. 
We use shared weights for the input and output embedding tables to reduce the total number of parameters~\citep{output-embeddings, attention-is-all-you-need}.

\subsection{Model training}

All training hyperparameters are the same for both models, GPT-2 and RETRO. 
For optimization, we use the Adam algorithm~\citep{adam-optimizer-kingma2017adam} with $\beta_1 = 0.9$ and $\beta_2 = 0.95$. 
The learning rate is set to $6\times10^{-4}$ and is reduced using a cosine scheduler by a factor of 10 by the end of training. 
For regularization, we use dropout layers with a probability of $0.1$, and the weight decay factor is $0.01$.
Table~\ref{tab:number-iterations} summarizes the number of training iterations.
For comparing models of all sizes (6, 9, and 12 layers), we conduct training on a smaller number of iterations.
For comparing the final models of sizes 6 and 9 layers, we conduct training on the larger number of iterations. 

\begin{table}[h]
\centering
\begin{tabular}{@{}lrrr@{}}
\toprule
Tokenization & Batch size & \# Iterations & \# Tokens \\
\midrule
GPT-2 & 384 & \num{20312} & $3.0\times 10^9$ \\
GPT-2 & 496 & \num{379032} & $72.2\times 10^9$ \\
%\hline
StarCoder & 384& \num{15234} & $2.2\times10^9$ \\
StarCoder & 512& \num{367187} & $72.2\times 10^9$ \\
\bottomrule
\end{tabular}
\caption{The number of training iterations. Training on eight A100 40 GB GPUs took 2 hours for the smaller number of iterations and 38 hours for the larger number of iterations.}
\label{tab:number-iterations}
\end{table}

\paragraph{Implementation details}
Our implementation of RETRO and GPT-2 follows the open-source implementation of both models by~\citet{retro++-wang2023shall}, which is in turn based on the original work by~\citet{retro-borgeaud22}.
For model training, we use the Megatron-LM framework, which is optimized for training LLMs on multiple GPUs that can be distributed across multiple compute nodes~\citep{megatronlm}.
In our case, all model training and data processing are conducted on a compute node with eight A100 40 GB GPUs.
Model training and execution on GPUs is performed using the \emph{bf16} floating-point format. 
% and the \emph{FlashAttention} algorithm for computing self-attention~\citep{dao2022flashattention}.

\subsection{Metrics}\label{sec:metrics}

For evaluating LLMs, we use different metrics for single-token and multi-token predictions. 
Single-token metrics are used to monitor model training because they are simple and efficient to compute. 
To assess multi-token predictions, we compare the generated output with the target output.

\subsubsection{Single-token metrics}
During model training, we use perplexity ($\PPL$), recall-at-k ($\Rk$) and mean reciprocal rank ($\MRR_k$) metrics~\citep{dive-into-deep-learning-book-zhang2023dive, information-retrieval-Jurafsky:2009:SLP:1214993}.
While $\PPL$ is a widespread metric, we find that recall-based metrics provide more intuitive interpretations of results compared to $\PPL$.

Let $x = (x_1, ..., x_n) \in \vocab^n$ be a sequence of tokens.
Perplexity $\PPL \in [1, \infty)$ measures the model's uncertainty about the data and is defined as
\begin{equation*}
    \PPL(x) = \exp\{-\frac{1}{n} \sum_{t=1}^{n}\log \prob_\theta \left(x_t \mid x_{t-1}, ..., x_1\right) \},
\end{equation*}
where $\log \prob_\theta \left(x_t \mid x_{t-1}, ..., x_1\right)$ is the log-probability for token $x_t$, as predicted by the model with parameters $\theta$.
We aim to minimize $\PPL$ to its optimal value of 1, which is achieved when the model correctly predicts each token with a probability of 1.

For easier interpretation, we also use recall metrics that summarize the probability that the predicted token is one of the correct tokens. 
Let $\hat{y}_i \in \R^{|\vocab|}$ be the output distribution of the next token after $x_i$, sorted by descending probability (i.e., the first element of $\hat{y}_i$ corresponds to the most probable next token). 
Then, $\rank_i$ is the index in $\hat{y}_i$ that corresponds to the next token $x_{i+1}$.

The recall-at-$k$ metric, $\Rk \in [0, 1]$, equals 1 for $x_i$ if $x_{i+1}$ is among the $k$ most probable suggested tokens.
For the entire $x$, we take the average
\begin{equation*}
    \Rk = \frac{1}{n} \sum_{i=1}^{n} \indicator \left( \rank_i \leq k \right).
\end{equation*}
$\Rk$ can be interpreted as the probability that the next token is among the $k$ most probable predicted tokens.

The Mean Reciprocal Rank $\MRR_k \in [0, 1]$ is defined as
\begin{equation*}
\MRR_k = \frac{1}{n} \sum_{i=1}^n \frac{1}{\rank_i},
\end{equation*}
where $\frac{1}{\rank_i}$ equals 0 if $\rank_i > k$.
We aim to maximize $\Rk$ and $\MRR_k$ to their optimal value of 1.
% The metrics $\Rena$, $\Rpet$, and $\MRRpet$ are reported in percentages.
When comparing single-token metric results, we must be cautious when comparing results obtained with different tokenizations, as the size and content of the vocabularies affect the obtained values.

\subsubsection{Multi-token predictions}
For multi-token predictions, we evaluate the quality of code completion, not only based on the prediction of the next token, but on generating multiple tokens.
We calculate metrics on \emph{strings}, removing leading and trailing whitespace from the strings~\citep{ReACC-lu2022reacc, repocoder-zhang-etal-2023-repocoder}.

Let $s$ be the model's prediction and $t$ the target string.
Exact Match $\exactmatch(s,t) \in \{0, 1\}$ is 1 if the strings $s$ and $t$ match exactly and 0 otherwise.
Edit Similarity $\editsim(s,t) \in [0, 1]$ considers how many edits are needed to transform $s$ into $t$:
\begin{equation*}
\editsim(s,t) = 1 - \frac{\lev(s, t)}{\max(|s|, |t|)},
\end{equation*}
where $\lev$ is the Levenshtein distance~\citep{levenshtein1966binary}.
When combining multiple pairs $(s,t)$ from a sample of predicted and target results $\mathcal{Y}$ for $\exactmatch$ and $\editsim$, we calculate the average, $\exactmatch = \frac{1}{n}\sum_{(s,t)\in\mathcal{Y}}\exactmatch(s,t)$ and similarly $\editsim = \frac{1}{n}\sum_{(s,t)\in\mathcal{Y}}\editsim(s,t)$.

Prefix Similarity $\prefixsim(s,t) \in [0, 1]$ measures the prefix matching of characters in the strings $s$ and $t$.
For $\prefixsim \in [0, 1]$, we take into account that lines of code have different lengths when combining multiple examples:
\begin{equation*}
\prefixsim = \frac{\sum_{(s,t)\in\mathcal{Y}}|\pi(s,t)|}{\sum_{(\cdot, t) \in \mathcal{Y}} |t|},
\end{equation*}
where $\pi(s,t)$ denotes the longest common prefix of $s$ and $t$.
We aim to maximize all three metrics $\exactmatch$, $\editsim$, and $\prefixsim$ to their optimal value of 1.

When reporting multi-token prediction metrics, we calculate 95\% confidence intervals obtained using the BCa~\citep{BCa-bootstrap-doi:10.1080/01621459.1987.10478410} bootstrap method with 1000 samples.
The notation \CI{a}{b}{c} means that $a$ is the calculated metric on the sample, and $b$ and $c$ are the endpoints of the confidence interval $[b,c]$.
% Metrics are reported in percentages.

\section{Results}\label{sec:results}
In this section, we report the results of experiments outlined in Section \ref{sec:experimental-setup}. We start by comparing the performance of RETRO and GPT-2 on the StarCoder dataset, followed by the evaluation on the RepoEval dataset, which simulates retrieval from local projects. In this setting, we also compare the In-context RAG approach. We finish the section with a qualitative analysis of the results.

\subsection{Evaluation on the StarCoder test set}\label{sec:test-results}
In this section, we present the results of our trained models on the test set from Section~\ref{sec:dataset}.
We are interested in the contribution of retrieval with RETRO compared to the baseline GPT-2 model.
We first examine the impact of the retrieval database size, tokenization, and the number of parameters on model performance. We also compare retrieval from the training and test set simulating different numbers of projects in a local database.

%\subsubsection{Test set results}
First, we examine the results of single-token metrics on the test set.
When reporting metrics for RETRO, the retrieval database is built from the union of the training and test sets, while during training only the training set is used~\citep{retro-borgeaud22, retro++-wang2023shall}.
If retrieval from the test set was forbidden, RETRO would not be able to find related code snippets from the same project as the test example, due to the project-level split into training and test sets.
We also report metrics calculated starting from the end of the first RETRO chunk $\offset{x} = (x_{64}, ..., x_n)$ instead of the entire input sequence $x = (x_1, ..., x_n)$.
For the first $m - 1 = 63$ tokens, RETRO does not use retrieval and therefore functions like the baseline model.
The notation \gptSgpt\ refers to the GPT-2 model with 6 layers and GPT-2 tokenization, while \gptSsc\ refers to the using the model with StarCoder tokenization (and similarly \retroSgpt\ for RETRO). 

\begin{figure*}[h]
    \centering
     \includegraphics[width=0.495\textwidth]{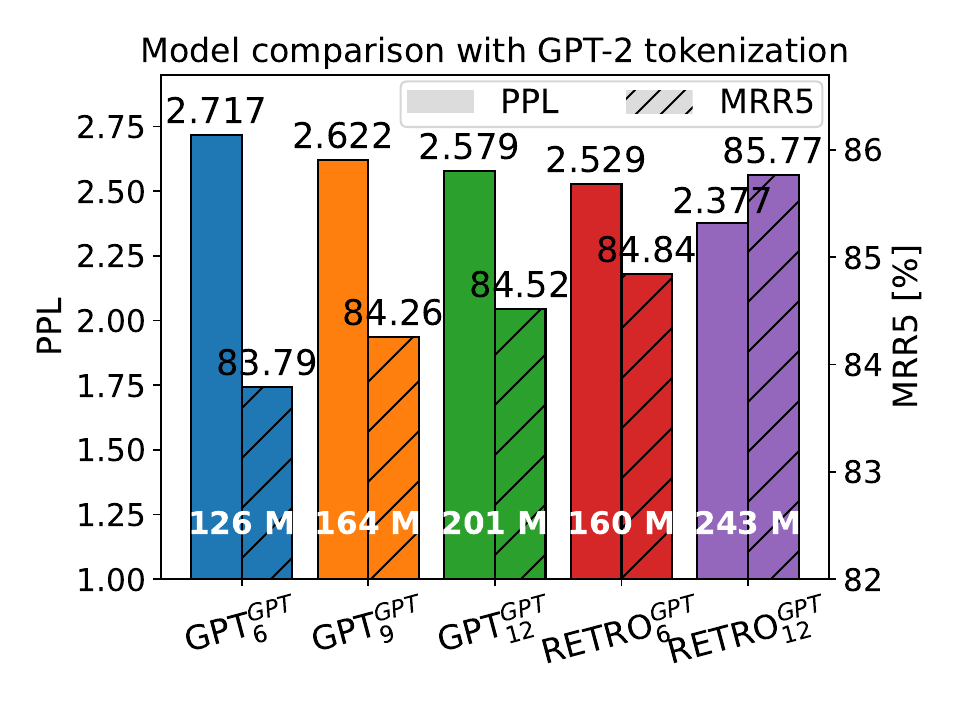}
     \includegraphics[width=0.495\textwidth]{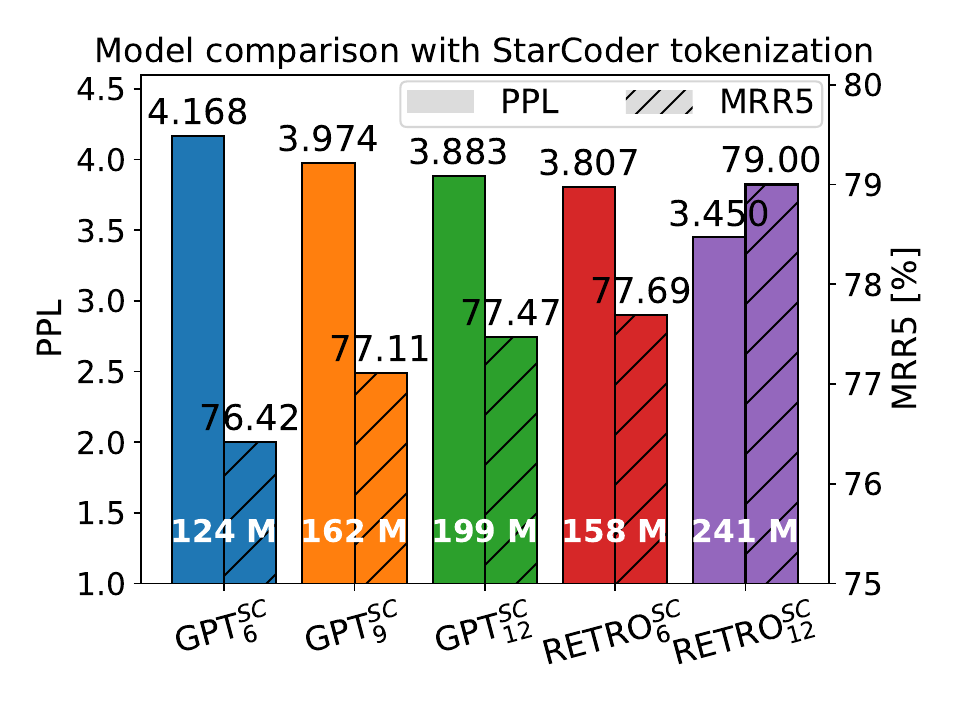}
    \caption{
  $\PPL$ and $\MRRpet$ metrics on the test set with models trained on a \textbf{smaller} number of iterations (see Table~\ref{tab:number-iterations}) with \emph{GPT-2} tokenization on the left and \emph{StarCoder} tokenization on the right.
  For each model, the left bar shows $\PPL$ and the right bar shows $\MRRpet$.
  Reported metrics are calculated on the offset input $\offset{x}$.
  The values of the left and right results should not be directly compared due to different tokenizations.
  The figures also show the models' sizes.
    }
    \label{fig:single-token-small-metrics}
\end{figure*}

\begin{table*}[h]
  \centering
  \renewcommand{\arraystretch}{1.15}
  %\small
  \begin{tabular}{@{}lrrrrrrrr@{}}
    \toprule
    
    \multicolumn{1}{c}{} & \multicolumn{4}{c}{} & \multicolumn{4}{c}{Offset $\offset{x}$} \\
    \cmidrule{6-9}
    
    Model & $\PPL$ & $\Rena$ & $\Rpet$ & $\MRRpet$ & $\PPL$ & $\Rena$ & $\Rpet$ & $\MRRpet$ \\
    %\hline
    \midrule

    \gptSgpt & 2.708 & 79.27 & 90.25 & 83.74 & 2.445 & 80.97 & 91.41 & 85.25 \\
    %\hline
    
    \gptMgpt & 2.602 & 79.95 & 90.69 & 84.33 & 2.347 & 81.67 & 91.85 & 85.84 \\
    & \bolje{-3.91} & \bolje{0.85} & \bolje{0.49} & \bolje{0.70} & \bolje{-4.01} & \bolje{0.86} & \bolje{0.48} & \bolje{0.69} \\
    %\hline
    
    \retroSgpt & 2.536 & 80.40 & 91.11 & 84.78 & 2.257 & 82.23 & 92.46 & 86.50 \\
    & \textbf{\bolje{-6.35}} & \textbf{\bolje{1.43}} & \textbf{\bolje{0.95}} & \textbf{\bolje{1.24}} & \textbf{\bolje{-7.69}} & \textbf{\bolje{1.56}} & \textbf{\bolje{1.19}} & \textbf{\bolje{1.47}} \\

    %\hline\hline
    \midrule
    
    {GPT$_9^{\textsc{SC}}$} & 3.946 & 70.66 & 86.54 & 77.07 & 3.433 & 73.06 & 88.18 & 79.19 \\
    %\hline
    
    RETRO$_6^{\textsc{SC}}$ & 3.855 & 70.98 & 86.98 & 77.45 & 3.293 & 73.65 & 88.87 & 79.84 \\
    & \textbf{\bolje{-2.30}} & \textbf{\bolje{0.45}} & \textbf{\bolje{0.51}} & \textbf{\bolje{0.49}} & \textbf{\bolje{-4.08}} & \textbf{\bolje{0.81}} & \textbf{\bolje{0.78}} & \textbf{\bolje{0.82}} \\
    
    \bottomrule
    
  \end{tabular}
  \caption{
  Single-token metrics on the test set with models trained on a \textbf{larger} number of iterations (see Table~\ref{tab:number-iterations}).
  Reported are also the relative differences of metrics compared to the baseline GPT-2 models.
  }
  \label{tab:single-metrics}
\end{table*}

In Figure~\ref{fig:single-token-small-metrics} we compare our models of all sizes trained on a smaller number of iterations and in Table~\ref{tab:single-metrics} we report the single-token metrics of our smaller sized models trained over more iterations.
In all cases we observe that the single-token metrics for RETRO are better than those for GPT-2 of comparable size.
However, the difference between the two models decreases when comparing models with approximately equal number of parameters, i.e. RETRO with 6 layers to GPT-2 with 9 layers, as seen in Table~\ref{tab:single-metrics}.
In Table~\ref{tab:single-metrics} we observe that the differences between RETRO and GPT-2 are higher when calculating metrics from token 64 onwards, suggesting the usefulness of the retrieval mechanism.
Increasing the number of layers and thus the number of model parameters also contributes to improvements in metrics, as is evident from Figure~\ref{fig:single-token-small-metrics}.
However, by training smaller models over more iterations, we achieve better metrics than using larger models trained for fewer iterations.
% The phenomenon of undertrained models is common in practice~\citep{chinchilla-hoffmann2022training}.
% It is recommended to double the number of training tokens each time the number of model parameters is doubled.
% Consequently, we did not train larger models for a longer number of iterations and omit them from Table~\ref{tab:single-metrike-ucenja}.
The relative differences are larger in bigger models, which may be due to the use of more $\retrolayer$ layers, as well as the effect of 21\% more parameters in \retroLgpt\ compared to \gptLgpt. 

% Differences in single-token metrics between GPT-2 and StarCoder tokenization are due to the different vocabularies. % \mrs{For such statements, we shall refer to supporting tables.}
We hypothesize that the significant differences in single-token metrics between GPT-2 and StarCoder tokenization are due to the different vocabularies making direct comparison of results less meaningful, so we should instead compare relative differences.
Nevertheless, we surmise that GPT-2 tokenization achieves better single-token metrics in Figure~\ref{fig:single-token-small-metrics} and Table~\ref{tab:single-metrics} due to the whitespace issue from Table~\ref{tab:tokenization-example}, as the model gets \enquote{easy points} by predicting individual spaces.
Despite larger absolute differences in individual metrics between GPT-2 and StarCoder tokenization, the relative differences in Figure~\ref{fig:single-token-small-metrics} and in Table~\ref{tab:single-metrics} are of comparable magnitude, suggesting a similar contribution of RETRO regardless of the tokenization used.
% In Table~\ref{tab:single-metrics}, the relative differences of the \retroSsc\ model seem smaller than in Table~\ref{tab:single-metrics-small-starcoder} because the differences in Table~\ref{tab:single-metrics} are reported against \gptMsc\ instead of \gptSsc.
% For a better comparison of models, it is necessary to perform comparisons of multi-token predictions, which we do in Section~\ref{sec:repoeval-evaluation}.

\paragraph{Projects in a local database}\label{sec:singletoken-projects}
For practical local use of RETRO, a small number of parameters would not be beneficial if we needed a retrieval database with a billion tokens to see benefits.
Therefore, we examine the contribution of RETRO using retrieval databases constructed from \emph{individual} projects in the test set by sampling 500 projects, such that the projects are evenly distributed with respect to the number of files, up to a maximum of 300 files.

\begin{figure*}[h]
\centering
\includegraphics[width=0.73\textwidth]{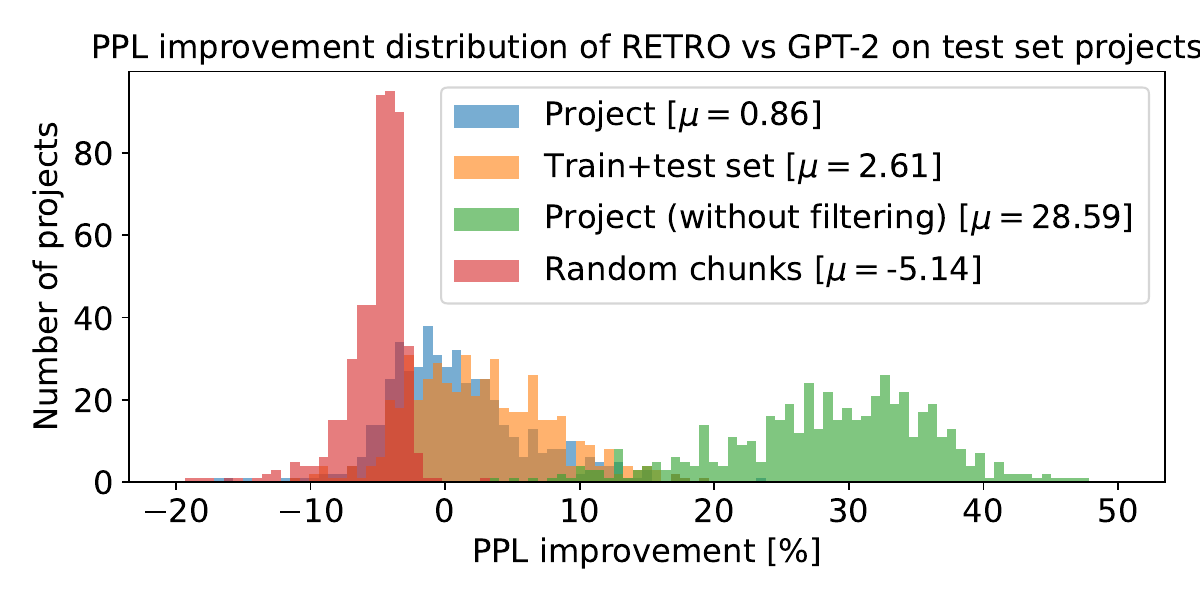}
\caption{Distribution of relative improvement in $\PPL$ between  \retroSgpt\ and \gptMgpt\ on 500 projects from our StarCoder test set based on the used retrieval database.
% Using random contexts serves as a lower bound, while using projects without filtering serves as an upper bound for RETRO's capabilities.
 }
\label{fig:ppl-distribution-singlerepo}
\end{figure*}

In Figure~\ref{fig:ppl-distribution-singlerepo}, we compare the relative improvement of \retroSgpt\ against \gptMgpt\ on projects using different retrieval databases. 
We limit the comparison to RETRO with 6 layers and GPT-2 with 9 layers to reduce the impact of the difference in the model sizes. 
Retrieving from the entire StarCoder training set contributes to a greater improvement compared to retrieving only from individual projects, which is consistent with the findings of \citet{retro-borgeaud22}, who observe improvements with increasing the retrieval database size. % up to $1.75$ trillion tokens. 
% From the shape of the distribution, we observe that retrieving only from projects often does not lead to improvement, but there are instances where retrieving is very beneficial (smaller local peak at a 10\% improvement), resulting in a positive contribution on average. 
% On the other hand, the contribution from retrieving from the entire StarCoder set is more frequent, with fewer cases of regression, indicating the benefits of a larger retrieval database.
We also test retrieval from the project without removing snippets originating from the same file as the input context, which gives an upper bound estimate for RETRO since the found snippets contain the correct next tokens. 
As expected, the difference with GPT-2 is in this case more significant, but it does not reach a perfect success rate. 
A lower bound estimate for RETRO is obtained when we retrieve random chunks from the entire training set.
In this case, we observe that the 6-layer \retroSgpt\ performs worse than the 9-layer \gptMgpt, which is expected, as it does not benefit from retrieval and consequently behaves as a 6-layer GPT-2 with added noise.

From Table~\ref{tab:singlerepo-summary}, we see that the single-token metrics using only \retroSgpt\ with the training set (without the test set) for retrieval are comparable to those of \gptMgpt.
This can be attributed to the fact that retrieving from the training set does not provide new information, as both models have seen the entire training set during training.
We therefore conclude that retrieval is only meaningful on new, previously unseen data for the model.
When retrieving from projects, the data is new, as the projects are sampled from the test set.
An implication of this observation is that it might be beneficial to train RETRO using a retrieval database different from the training set.
The training could also include examples of random chunks to help the model better learn to skip irrelevant information, and examples of same-document snippets to help the model better learn to copy correct answers.

\begin{table*}[htb]
  \centering
  \renewcommand{\arraystretch}{1.3}
  \begin{tabular}{@{}lrrrr@{}}
    \toprule
      Retrieval database & $\Delta\PPL$ & $\Delta\Rena$ & $\Delta\Rpet$ & $\Delta\MRRpet$ \\
      % Retrieval database & $\Delta\PPLoff$ & $\Delta\Renaoff$ & $\Delta\Rpetoff$ & $\Delta\MRRpetoff$ \\
    \midrule

    % \gptMgpt & \CI{2.49}{2.42}{2.57} & \CI{81.01}{80.56}{81.40} & \CI{91.88}{91.58}{92.11} & \CI{85.48}{85.09}{85.80} \\
    % \hline\hline
    
    Project & \CI{0.86}{0.42}{1.36} & \CI{0.12}{0.01}{0.23} & \CI{0.24}{0.17}{0.31} & \CI{0.18}{0.09}{0.27} \\
    %\hline
    
    Training set & \CI{-0.05}{-0.45}{+0.37} & \CI{-0.03}{-0.12}{+0.07} & \CI{0.12}{0.06}{0.18} & \CI{0.05}{-0.03}{+0.13} \\
    %\hline
    
    Training + test set & \CI{\textbf{2.61}}{2.15}{3.04} & \CI{\textbf{0.53}}{0.43}{0.64} & \CI{\textbf{0.51}}{0.44}{0.58} & \CI{\textbf{0.54}}{0.45}{0.63} \\
    %\hline\hline
    
    Project \emph{(without filtering)} & \CI{28.59}{27.89}{29.39} & \CI{7.19}{6.99}{7.44} & \CI{4.78}{4.64}{4.97} & \CI{6.24}{6.06}{6.45} \\
    %\hline
    
    Random chunks & \CI{-5.14}{-5.39}{-4.94} & \CI{-1.10}{-1.17}{-1.05} & \CI{-0.61}{-0.64}{-0.58} & \CI{-0.89}{-0.94}{-0.85} \\
    \bottomrule

  \end{tabular}
  \caption{
  Comparison of the relative improvement of single-token metrics between \retroSgpt\ in \gptMgpt\ on 500 projects from the test set with respect to the used retrieval database.
  The metrics are in percentages, contain 95-percent confidence intervals and are calculated on the offset input $\offset{x}$.
  }
  \label{tab:singlerepo-summary}
\end{table*}

\subsection{Completing code lines from local projects with RepoEval}\label{sec:repoeval-evaluation}
% Similar to Section~\ref{sec:singletoken-projects}, we are interested in the practical use of models on local projects, but this time we are considering multi-token predictions.
In this section, we aim to simulate and evaluate completing lines in local projects using the RepoEval dataset, as described in Section~\ref{sec:repoeval-dataset}.
% The input context for models without retrieval contains only the lines before the target line within a single file. %, while retrieval includes useful information from other files in the project.
% With retrieval, the retrieval database is built only from individual projects (we do not use the entire StarCoder training set).
% In our experiments, all text is generated using greedy decoding to achieve determinism and faster execution when completing short lines of code.
We first present the results of RETRO, GPT-2, and In-context RAG in this setting, followed by an experiment with larger models, and detailed analysis of our most successful approach, namely In-context RAG. We end the section with an analysis of the impact the quality of the retrieved snippets has on the code-completion performance.

\subsubsection{RETRO and GPT-2 results on local projects}

\begin{figure*}[htb]
    \centering
    \includegraphics[width=0.495\linewidth]{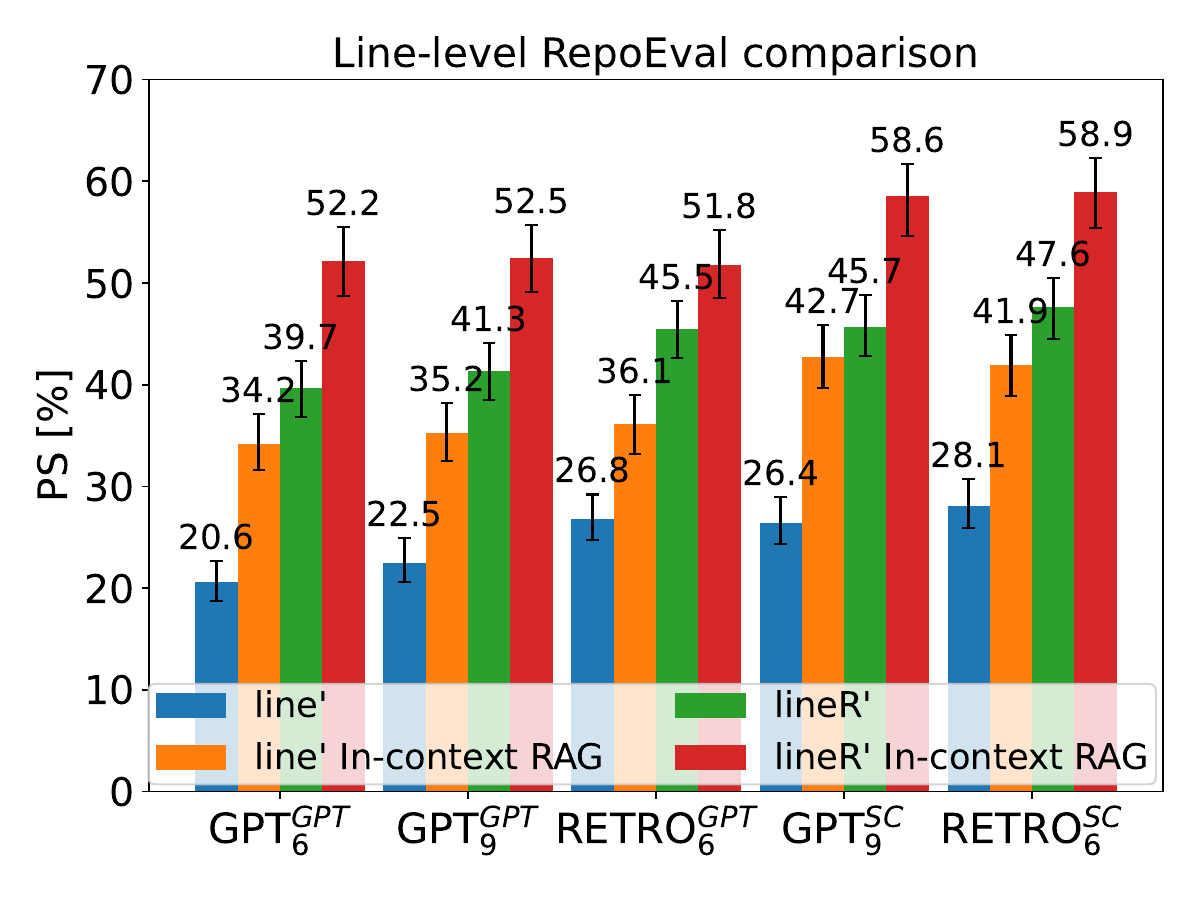}
    \includegraphics[width=0.495\linewidth]{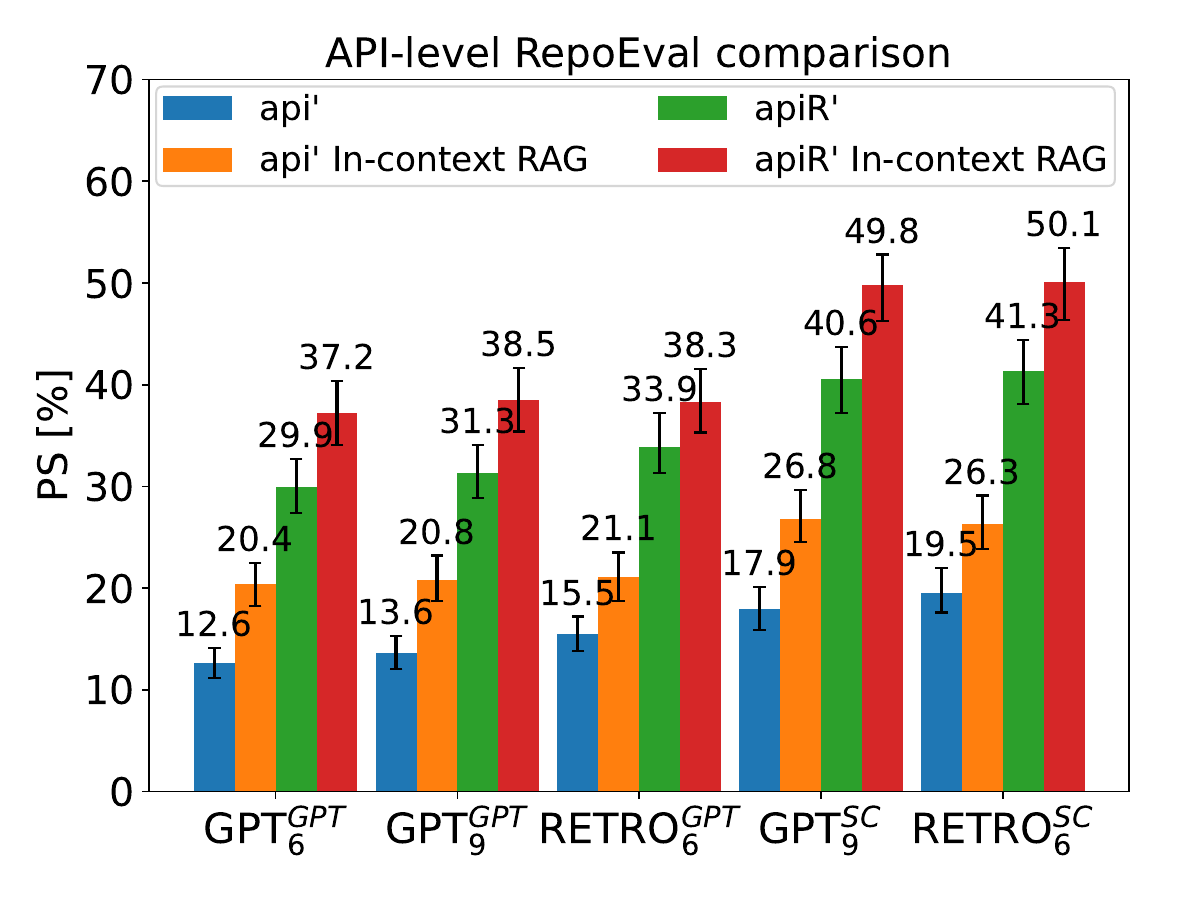}
    \caption{
    Comparison of our models using GPT-2 and StarCoder tokenization on the RepoEval line-level (left) and API-level (right) datasets without overlap with the training set.
    The bar plots contain 95-percent confidence intervals of the prefix similarity metric.
    }
    \label{fig:repoeval-small}
\end{figure*}

In Figure~\ref{fig:repoeval-small}, we compare our trained models, RETRO and GPT-2, using both GPT-2 and StarCoder tokenizations.
The metrics confirm the advantage of StarCoder tokenization, which better handles the whitespace issue mentioned in Section~\ref{sec:tokenization}.
Similar to the single-token metrics, with multi-token predictions, we observe that the 6-layer RETRO achieves significantly better performance than the 6-layer GPT-2.
However, the difference between the 9-layer GPT-2 and the 6-layer RETRO is not as pronounced, especially with the StarCoder tokenization.

%%% Option 1 - Two figures %%%

Metrics on \repoevallinerand\ are higher than on \repoevalline\ (similarly for \repoevalapirand\ and \repoevalapi), because in the case of \repoevalline\ we predict entire lines, while in \repoevallinerand\ the model already has some additional context about the content of the current line.
This is important because predicting the entire line requires correctly predicting the start of the line; otherwise, the $\exactmatch$ and $\prefixsim$ metrics are immediately zero.
Predicting the correct start of a line is not always straightforward, as multiple beginnings can make sense. % (e.g., the next line could be a comment, a print statement, or the introduction of a variable with a different name).

In Figure~\ref{fig:repoeval-small}, we also report the metrics for the In-context RAG approach (more in Section~\ref{sec:rag-results}). 
Despite its simplicity, the contribution of the In-context RAG is more significant than the use of RETRO.
When combining In-context RAG with RETRO, we observe smaller improvements compared to enhancing GPT-2 with In-context RAG.
A possible explanation is that including retrieved snippets into the input context \enquote{confuses} RETRO's chunk based retrieval, which now finds neighbors of the already retrieved snippets for part of the context.
Additionally, snippets included in the context receive the same treatment as other input tokens through processing in the transformer, while RETRO incorporates the retrieved chunks using cross-attention only in the sparsely dispersed \retrolayer\ layers.
% The advantages of the RETRO architecture and In-context RAG could be better combined with additional fine-tuning of the model using examples that include retrieved snippets in the input context~\citep{retro++-wang2023shall}.

\subsubsection{Comparison with larger models}

%% line + api with PS
\begin{table*}[htb]
  \centering
  \renewcommand{\arraystretch}{1.25}
  %\footnotesize
  \begin{tabular}{@{}llrrrrrr@{}}
    \toprule
    \multicolumn{2}{c}{} & \multicolumn{3}{c}{} & \multicolumn{3}{c}{In-context RAG} \\
    \cmidrule{6-8}
    %\hline
    Model & Data & $\exactmatch$ & $\editsim$ & $\prefixsim$ & $\exactmatch$ & $\editsim$ & $\prefixsim$  \\
      \midrule

      \gptMsc & {\repoevalline} & \CI{18.0}{16.1}{19.9} & \CI{47.0}{45.3}{48.4} & \CI{23.2}{21.4}{25.0} & \CI{33.9}{31.6}{36.0} & \CI{58.4}{56.7}{60.2} & \CI{40.9}{38.4}{43.6} \\
      & {\repoevalapi} & \CI{8.8}{7.3}{10.1} & \CI{39.3}{37.8}{40.6} & \CI{15.0}{13.5}{16.3} & \CI{21.1}{19.1}{23.2} & \CI{51.7}{50.0}{53.3} & \CI{26.9}{24.9}{29.0} \\
      %\hline
      
      \retroSsc & {\repoevalline} & \CI{19.6}{17.7}{21.8} & \CI{50.1}{48.6}{51.8} & \CI{26.4}{24.5}{28.1} & \CI{31.6}{29.2}{33.8} & \CI{57.2}{55.6}{58.9} & \CI{39.5}{37.3}{42.2} \\
      & {\repoevalapi} & \CI{9.6}{8.1}{11.0} & \CI{42.7}{41.3}{44.1} & \CI{16.8}{15.4}{18.2} & \CI{20.1}{18.1}{21.9} & \CI{50.5}{48.6}{51.9} & \CI{26.3}{24.4}{28.3} \\
      %\hline
      
      \tinystarcoder & {\repoevalline} & \CI{25.1}{22.9}{27.0} & \CI{53.1}{51.4}{54.7} & \CI{31.9}{29.9}{34.2} & \CI{41.6}{39.1}{44.0} & \CI{64.6}{62.6}{66.3} & \CI{48.1}{45.2}{50.9} \\
      & {\repoevalapi} & \CI{23.2}{21.0}{25.1} & \CI{52.6}{50.9}{54.1} & \CI{32.4}{30.2}{34.6} & \CI{34.5}{32.2}{36.8} & \CI{63.0}{61.4}{64.6} & \CI{43.6}{41.4}{46.0}  \\
      %\hline
      
      \bigstarcoder & {\repoevalline} & \CI{\textbf{39.4}}{37.1}{41.8} & \CI{\textbf{64.4}}{62.7}{66.2} & \CI{\textbf{44.8}}{42.3}{47.5} & \CI{51.0}{48.8}{53.4} & \CI{72.2}{70.6}{73.7} & \CI{\textbf{56.5}}{53.8}{59.2} \\
      & {\repoevalapi} & \CI{32.1}{29.8}{34.2} & \CI{61.2}{59.4}{62.9} & \CI{41.3}{39.0}{43.8} & \CI{44.1}{41.5}{46.4} & \CI{71.4}{69.8}{72.9} & \CI{53.6}{51.2}{56.1} \\
      %\hline
      
      \davinci & {\repoevalline} & 38.69 & 62.58 & / & \textbf{55.94} & \textbf{74.34} & / \\
      & {\repoevalapi} & \textbf{32.50} & \textbf{61.52} & / & \textbf{47.75} & \textbf{73.30} & / \\ 
      \bottomrule
  \end{tabular}
  \caption{
  Results of comparing smaller and larger models on RepoEval.
  The results for \davinci\ are from~\citet{repocoder-zhang-etal-2023-repocoder}.}
  \label{tab:repoeval-bigmodels}
\end{table*}

\begin{figure}[h]
    \centering
    \includegraphics[width=0.56\linewidth]{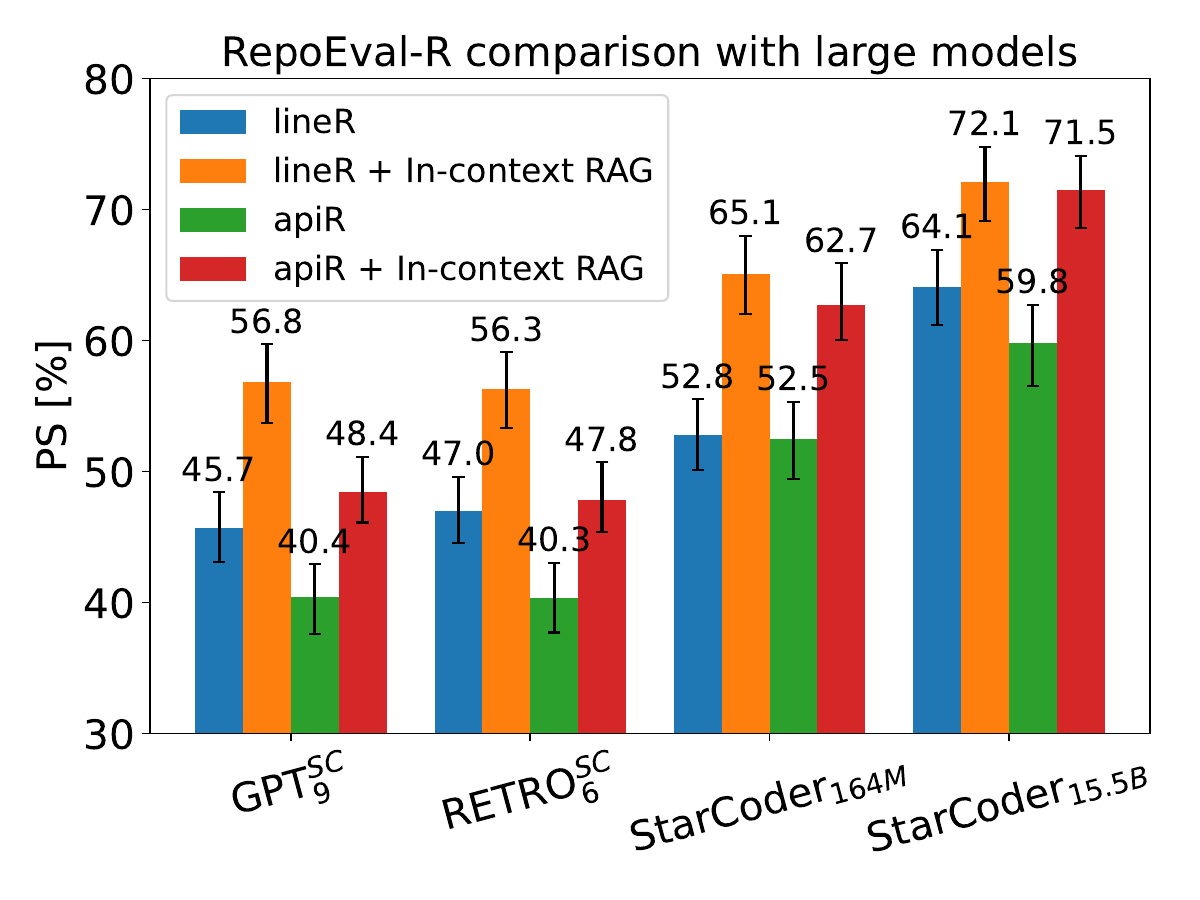}
    \caption{
    Comparison of our models with larger models on the line-level and API-level RepoEval datasets starting completion from random positions.
    The bar plots contain 95-percent confidence intervals of the prefix similarity metric.
    }
    \label{fig:repoeval-big-models-rand}
\end{figure}

Using small models is beneficial for local model execution, but one of the advantages of LLMs is their ability to scale by increasing the number of parameters~\citep{kaplan2020scaling}. 
In Figure~\ref{fig:repoeval-big-models-rand} and Table~\ref{tab:repoeval-bigmodels}, we examine the contribution of larger models compared to our smaller models.
% We also include results from~\citet{repocoder-zhang-etal-2023-repocoder}, where they tested the commercial \davinci\ model. 
The open-source models \tinystarcoder\ (with $164\times10^6$ parameters) and \bigstarcoder\ (with $15.5\times10^9$ parameters) are trained on the entire StarCoder dataset and further fine-tuned on Python~\citep{li2023starcoder}.
\tinystarcoder\ serves as a good comparison to our models due to its comparable number of parameters and for assessing the contribution of training on a larger dataset with a larger context size (8192 compared to our size of 384).

%%% Option 1 - Small table + PS R-level figure %%% 

Increasing the number of parameters in some cases more than doubles the performance compared to smaller models, as seen in Table~\ref{tab:repoeval-bigmodels}. 
However, with the use of In-context RAG, even smaller models can achieve results comparable to larger models (without retrieval). 
The positive impact of retrieval is not limited to smaller models, as incorporating useful information from the project into the context benefits the completions of larger models as well.

Results on \repoevalapi\ are lower than on \repoevalline\ due to completing function calls from projects and also due to the longer length of examples, which can contain more than a single line. 
Consequently, the model must correctly predict more tokens, increasing the likelihood of error. 
For longer examples the differences are even greater due to the use of greedy decoding and the propagation of errors, which are more frequent in smaller models.

\subsubsection{In-context RAG}\label{sec:rag-results}
% In-context RAG is not limited to using RETRO, as the approach works with any autoregressive model. 
Since the found snippets are included directly in the model's context, it is important that the model supports a sufficiently large context to append the retrieved data in addition to the input context.
When using our models with a context size of 384 tokens, we often need to truncate the input sequence to leave enough space to include the retrieved snippets at the beginning of the context.
For our models, GPT-2 and RETRO, we include only one retrieved snippet due to the small context size of the model.
For models with larger context sizes, we include two retrieved snippets. 
In Table~\ref{tab:repoeval-bigmodels} and in Figures~\ref{fig:repoeval-small} and \ref{fig:repoeval-big-models-rand}, we report multi-token prediction metrics both with and without retrieval as described in Section~\ref{sec:incontext-jaccard}. 
The results indicate that using In-context RAG leads to improvements in multi-token predictions compared to the baseline models in all cases.

\paragraph{Copying}

\begin{figure}[tb]
    \centering
    \includegraphics[width=0.5\textwidth]{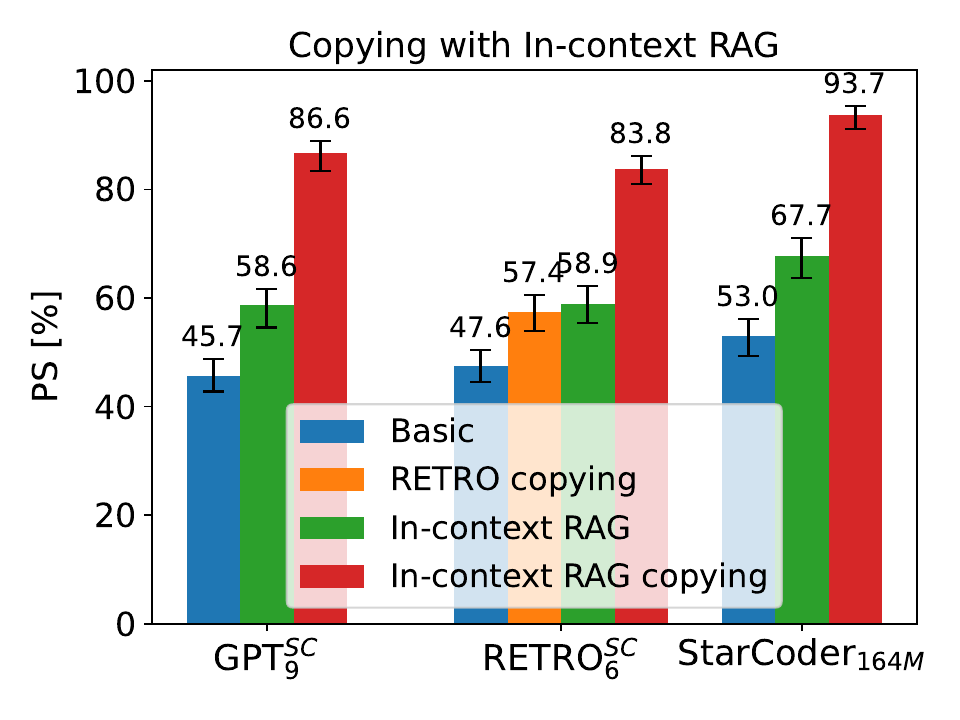}
    \caption{
    Comparison of copying abilities using In-context RAG on \linerandclean.
    With copying, retrieved snippets can be from the same file as the input context, which means including the correct continuation of the current line in the model's context.
    \textit{RETRO copying} shows the use of RETRO without filtering out the current file from retrieval but without additional In-context RAG (see also Section~\ref{sec:singletoken-projects}).
    The bars show 95\% confidence intervals for the prefix similarity metric $\prefixsim$.
    }
    \label{fig:copying}
\end{figure}

In Section~\ref{sec:singletoken-projects}, we estimate the upper bound of single-token metrics for RETRO by not removing chunks retrieved from the same file as the input context.
Similarly, we want to determine the upper bound of model performance when using In-context RAG if the retrieved snippets can contain the continuation of the current file we are completing.
This means that the context includes the line we want to complete, allowing the model to simply copy it from the context.
In Figure~\ref{fig:copying}, we see that copying from retrieved snippets is very effective, while copying with RETRO without In-context RAG performs similarly to using RETRO with In-context RAG without copying.

In addition to the explanations in Section~\ref{sec:singletoken-projects}, we also note that because the first \retrolayer\ layer is the sixth layer (which is also the last and only \retrolayer\ layer for \retroSsc), certain information from the input tokens may get diluted, preventing direct copying of tokens from the encoded chunks with chunked cross-attention.
We hypothesize that adding copying examples to the training would improve the model's ability to copy, and \retrolayer\ layers could be included closer to the beginning of processing instead of only in the sixth layer.
Accurate copying is important because when completing code, we want to extract not only the semantic meaning from the found snippets but also exact facts (e.g., the exact name of an existing function and not a hallucinated but semantically similar name).

\subsubsection{Impact of retrieval similarity}\label{sec:similarity-impact}

\begin{figure*}[htb]
  \centering
  \includegraphics[width=0.495\textwidth]{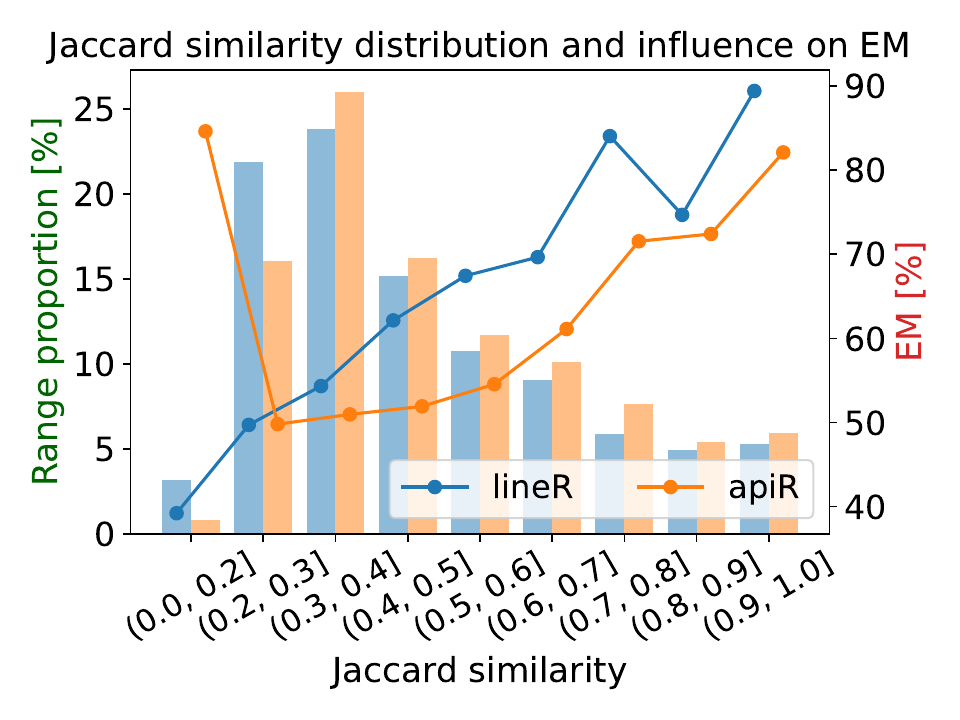}
  \includegraphics[width=0.495\textwidth]{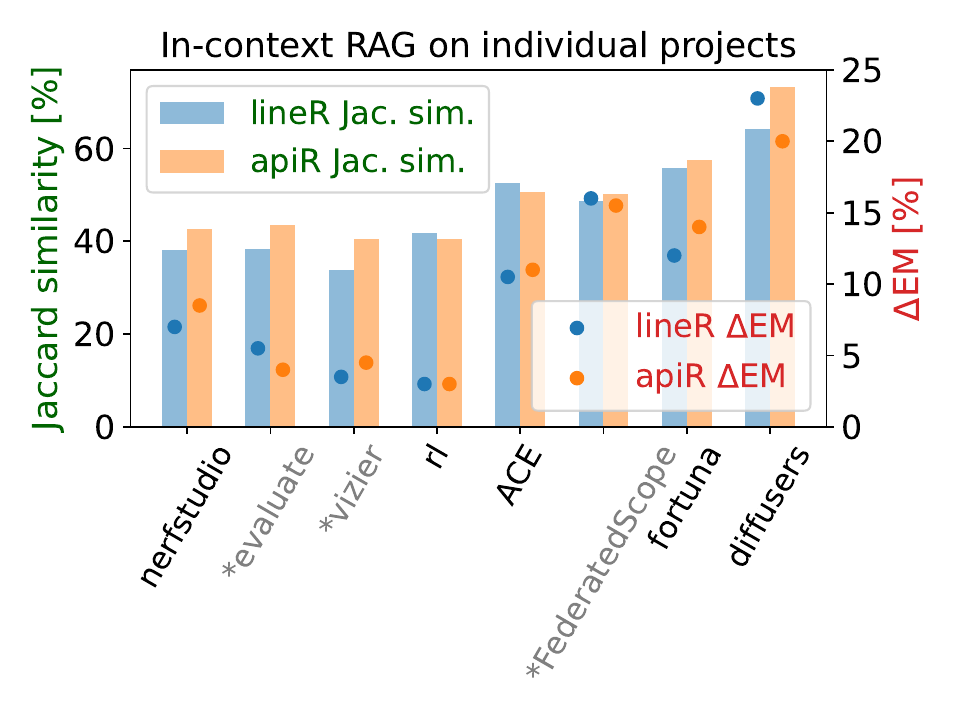}
  \caption{
  Impact of the Jaccard similarity coefficient evaluated on \repoevallinerand\ and \repoevalapirand\ using the \tinystarcoder\ model and In-context RAG. 
  The left side shows $\exactmatch$ increasing as the Jaccard similarity increases, and the distribution of similarities, which is slightly higher on \repoevalapirand\ than on \repoevallinerand.
  The right side shows the average Jaccard similarity for each of the eight projects in \repoevallinerand\ and \repoevalapirand\ and the difference in $\exactmatch$ for each project individually compared to the baseline model without retrieval.
  Projects with gray names and an asterisk are those removed in \linerandclean\ and \apirandclean\ due to a non-empty overlap with the training set.
  }
  \label{fig:jaccard-impact-repoeval}
\end{figure*}

Intuitively, we expect the benefit of retrieval-augmented line completion to depend on the quality of the retrieved snippets. 
Therefore, we first examine the impact of the retrieved snippets' similarities on the exact match metric $\exactmatch$ with the results shown in Figure~\ref{fig:jaccard-impact-repoeval}. 
As expected, the metric increases with higher matching according to the Jaccard similarity coefficient. 
We also find that most of the metric improvement comes from projects where snippets with high similarity to the query from the input context are found.
The benefits of retrieval are thus project-dependent, with greater benefits in projects with repetitive code, as a high Jaccard similarity indicates token overlap between files.

Using RETRO or In-context RAG generally improves multi-token predictions. 
However, there are cases where the model without retrieval correctly completes a line, while the model with retrieval does not, as summarized in Table~\ref{tab:rag-better-worse}. 
Most examples remain unchanged regardless of retrieval use, suggesting that retrieval is beneficial only in specific cases.
While positive examples dominate, the proportion of worsened results is not negligible and we want to reduce it. 

\begin{table*}[h]
  \centering
  \renewcommand{\arraystretch}{1.0}
  \begin{tabular}{@{}lcccc@{}}
    \toprule
    \multicolumn{2}{c}{} & \multicolumn{3}{c}{In-context RAG} \\
    \cmidrule{3-5} 
    & \retroSsc  & \retroSsc & \gptMsc & \tinystarcoder \\
    % & \retroSsc  & $_{\text{RAG}}$\retroSsc & $_{\text{RAG}}$\gptMsc & $_{\text{RAG}}$\tinystarcoder \\
    \midrule
    \gptMsc & \cmp{12.3}{7.9} & \cmp{20.3}{9.3} & \textbf{\cmp{17.4}{6.2}} & \cmp{26.2}{6.3} \\
    \retroSsc & / & \textbf{\cmp{14.5}{6.0}} & \cmp{17.2}{8.8} & \cmp{24.4}{6.6} \\
    \tinystarcoder & \cmp{11.4}{15.6} & \cmp{17.5}{14.6} & \cmp{17.2}{14.1} & \textbf{\cmp{15.8}{3.8}} \\
    \bottomrule
  \end{tabular}
  \caption{
  Proportions of improved and worsened examples based on prefix similarity when comparing the baseline model (rows) with the model with retrieval (columns) on the \repoevallinerand\ dataset.
  The first column of the first row tells us that \retroSsc\ completes the line better than \gptMsc\ in 12.3\% of cases, worse in 7.9\% of cases, and in the remaining cases, the predictions of both models are the same.
  }
  \label{tab:rag-better-worse}
\end{table*}

\begin{figure*}[h]
    \centering
    \includegraphics[width=0.495\textwidth]{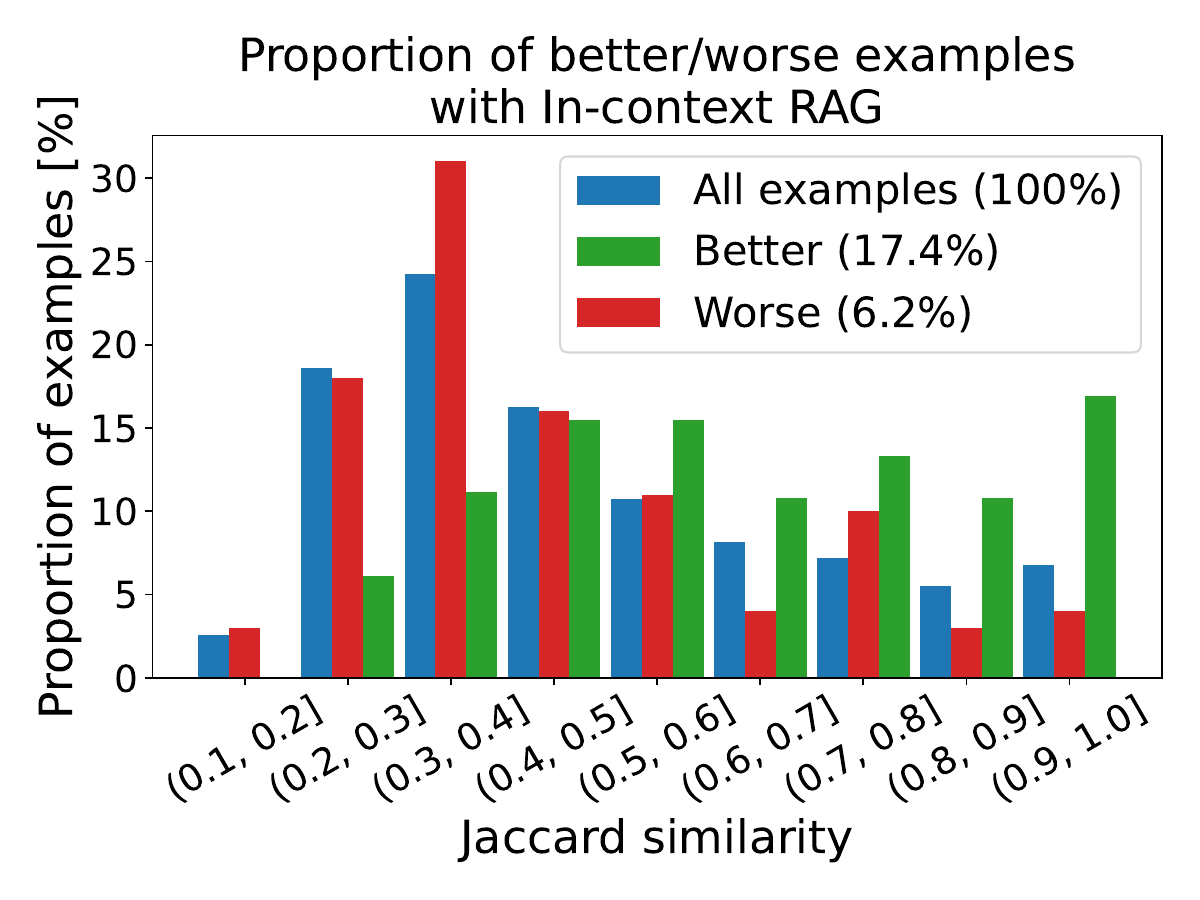}
    \includegraphics[width=0.495\textwidth]{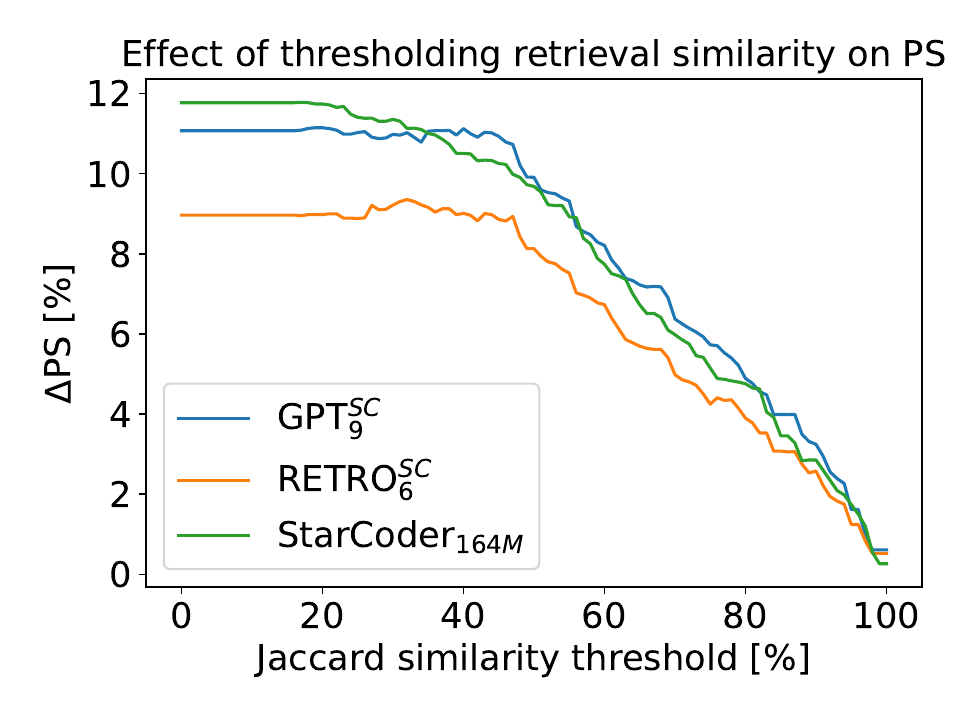}
    \caption{
    The left side shows the proportion of cases that are better or worse with In-context RAG compared to the baseline model without retrieval.
    The model used is \gptMsc\ on \repoevallinerand\ with $\prefixsim$ being compared.
    Retrieval is beneficial mainly when there is a high similarity of the retrieved snippets.
    The right side shows the difference in the $\prefixsim$ metric to the baseline model when limiting retrieval to cases above a certain similarity threshold.
    % At a 100\% similarity threshold, we get the baseline model without retrieval.
    }
    \label{fig:rag-thresholds}
\end{figure*}

Now we focus only on the examples highlighted in Table~\ref{tab:rag-better-worse}, where we detect worsening or an improvement in metrics.
From Figure~\ref{fig:rag-thresholds}, we find that the impact of retrieval is negative mainly at low similarities to the query, while the positive impact is more present at higher similarities.
To mitigate the consequences of poor retrieval, we examine the idea of whether it makes sense to limit retrieval to found snippets with similarities higher than some threshold value.
Unfortunately, the right side of Figure~\ref{fig:rag-thresholds} does not show a significant benefit from limiting to sufficiently similar snippets.

\subsection{Qualitative analysis}\label{sec:qualitative-analysis}
In Section~\ref{sec:repoeval-evaluation}, we found that retrieval generally benefits model predictions, but there are still cases where retrieval degrades the accuracy of predictions. 
In this section, we conduct a qualitative review of successful and less successful line completion examples, using our \gptMsc\ and \retroSsc\ models.

\begin{figure*}[h]
\centering
\includegraphics[width=0.6\textwidth]{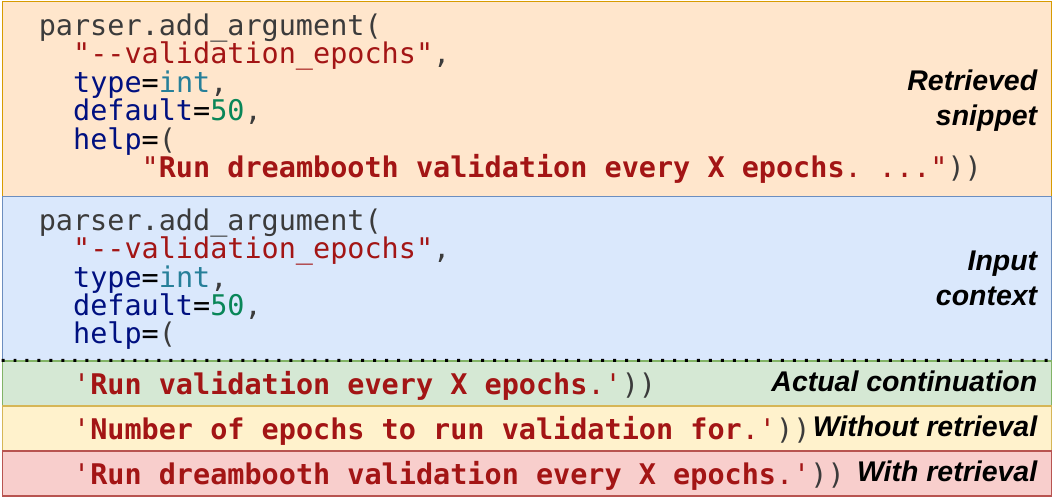}
\caption{
An example of completing documentation where retrieval guides the model to copy from a highly similar found snippet.
The code is shortened for clarity.
This example also shows the difficulty of evaluating natural language using string similarity based metrics, as the model's prediction without retrieval is very reasonable.
}
\label{fig:documentation}
\end{figure*}

\paragraph{Common errors}
Our small models (with or without retrieval) mostly make semantic errors; syntactic errors are rare. 
Semantic errors require a higher level of code understanding, which might exceed the capabilities of our small models. 
Difficulties often arise when completing natural language strings (e.g., comments and documentation within the code), at the beginning of files in \texttt{import} statements, and when using boolean or numerical constants. 
Degradation is also present in longer predictions due to greedy decoding and error propagation.
For completing \texttt{import} statements, the model would need information about the upcoming lines of the file to infer which modules to import into the program.
In completing documentation, we have a similar problem if we are documenting code that has not yet been written and follows in the subsequent lines.
Additionally, our string matching based evaluation does not distinguish between semantically equivalent ways of expressing messages causing examples like in Figure~\ref{fig:documentation} to achieve low metrics.
Nevertheless, retrieval can offer an advantage in such cases by providing the model with an example of the existing documentation style.
% The issue of the current line depending on subsequent lines is addressed by the \emph{fill-in-the-middle} technique~\citep{fim-donahue-etal-2020-enabling}, which includes lines that follow the current line into the context. 
% Including data from the same file is also possible by retrieving within the current file, but we avoid this to avoid leaking the target line.

\paragraph{Hallucinations}
Handling hallucinations in code completion is more manageable compared to natural language, as the proposed code can be verified with static code analysis tools to discard suggestions containing errors (e.g., calling a non-existent function or using an incorrect number of arguments). 
However, static analysis cannot resolve semantic errors, such as incorrect conditions in loops or the improper order of operations. 
Therefore, careful attention and understanding of the code are necessary when using code completion tools. 

\paragraph{Advantages and disadvantages of retrieval}

\begin{figure*}[hbt]
\centering
\includegraphics[width=1.0\textwidth]{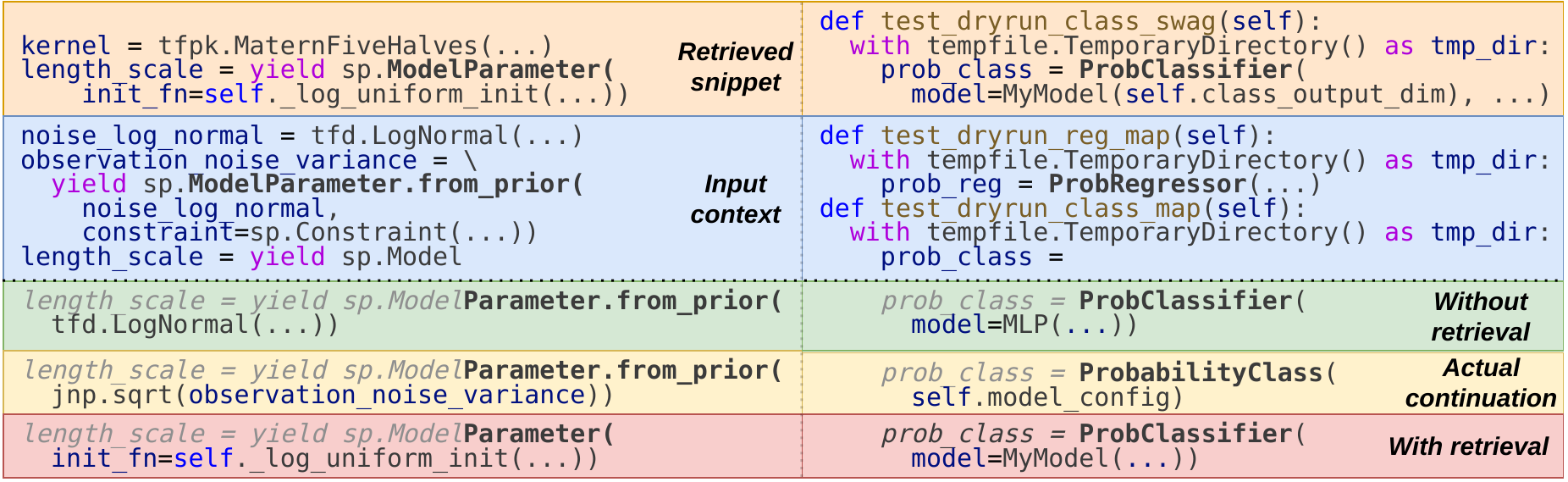}
\caption{
On the left is an example of misleading retrieval where the model incorrectly uses a function as it appears in the retrieved snippet instead of how it is used in the current file. 
On the right is an example of completing a test function in a project where a model with a longer context could correctly complete the lines, as the name \texttt{ProbClassifier} would have been encountered at the beginning of the file in the \texttt{import} statements. 
However, with a context size of 384, the beginning of the file is lost, but retrieval finds a similar example in other tests in the project, enabling successful completion compared to the baseline model. 
The code is shortened for clarity in both cases.
}
\label{fig:misleading}
\end{figure*}

As discussed in Section~\ref{sec:similarity-impact}, the advantage of retrieval-augmented completion becomes apparent when highly similar examples to the input context are found in the project. 
An example of repetitive code occurs when writing tests, which also involves the use of objects defined within the project. 
In most cases, the model can focus on the relevant information in the context and successfully skip over unhelpful retrieved snippets. 
However, when the model performs worse with retrieval, it is often due to misleading snippets that conflict with the information of the current file. 
For example, the model might use a function call as found in a related file instead of how it should be used according to the current file, as seen in Figure~\ref{fig:misleading}. 
% It also highlights the problem of models with a short context, such as our models with 384 tokens, when editing lines towards the end of the file, as the context no longer includes information from the beginning of the file. 
% In such cases, retrieval can still be beneficial by augmenting the context with important information.

\paragraph{Evaluation limitations}
Evaluating the quality of code completion using metrics is challenging~\citep{yaroslav-metrics-EVTIKHIEV2023111741}. 
In many cases, a line can be completed in multiple ways, but evaluation datasets typically contain only one correct answer. 
This problem could be mitigated by sampling multiple generated solutions, thereby increasing the likelihood of a correct match. 
A better solution would be to execute the generated code using unit tests defined in the project. 
However, besides the complexity of running code (e.g., setting up a development environment for each project), this is not the most suitable quality measure in our case since we are completing individual lines rather than entire functions. 
Instead of string based metrics, LLMs could be used to evaluate quality, as they have a higher correlation with human judgements~\citep{g-eval-gpt4-liu-etal-2023-g}. 
Ultimately, it is the developer who decides whether to accept the suggested line, so for evaluation, it would be necessary to ask users for feedback or conduct A/B tests of the final product integrated into a code editor.

% \paragraph{Summary of results}
% Comparing the single-token and multi-token results of our GPT-2 and RETRO models, we find that RETRO with retrieval from the local project improves metrics compared to both the 6-layer and the 9-layer GPT-2. 
% However, the difference between the 9-layer GPT-2 and the 6-layer RETRO is not pronounced, indicating the importance of considering the number of model parameters when comparing small models. 
% Using In-context RAG to retrieve from the local project enhances the accuracy of all evaluated models, both small and larger models, but the contribution depends on the quality of retrieval and the project's characteristics. 
% Despite the simplicity of the approach, In-context RAG matches the performance of GPT-2 with RETRO for line completion tasks. 
% By comparing copying abilities, we identify shortcomings in RETRO, which we suspect could be addressed primarily through changes in model training.
% Qualitative analysis identifies cases where retrieval is beneficial, but sometimes also misleads model predictions.
% We also highlight the shortcomings of our evaluation, which does not account for semantic similarity, which is more important when evaluating natural language completion and less so for short lines of code.

%\section{Discussion}\label{sec:discussion}

\section{Conclusion and future work}\label{sec:conclusion}

In this work, we examined the impact of retrieval from local projects on the success of completing lines of code using LLMs. 
We compared the GPT-2 and RETRO models, which we trained on a dataset of Python files from the StarCoder dataset. 
We compared two different tokenization approaches and highlighted the importance of token healing and the proper tokenization of whitespace when completing code, as it affects the amount of information in the context and the storage size of tokenized data.
% For training RETRO, we replaced the BERT encoder with a pretrained CodeT5+ encoder. 
% We sped up retrieval by indexing the embeddings and verified the indexing settings for efficient and high-quality retrieval. 
% We improved text generation with the trained models using token healing and chunk padding for RETRO.
We limited GPT-2 and RETRO to a smaller size of around 160 million parameters and reduced the impact of the differing number of parameters when comparing the two models.
We confirmed that RETRO with retrieval from the local project improves metrics compared to both the 6-layer and 9-layer GPT-2 on all considered benchmarks.
However, the difference between the 9-layer GPT-2 and the 6-layer RETRO is not pronounced, indicating the importance of considering the number of model parameters when comparing small models. 
% We estimated the upper and lower bounds of RETRO based on the retrieval database and examined the impact of the size of the retrieval database. 
We further examined the impact of retrieval using In-context RAG, which includes snippets found using the Jaccard similarity of tokens into the context.
Using In-context RAG to retrieve from projects in RepoEval improved results of both small and larger models, but the improvements depended on the quality of retrieval and the projects' characteristics. 
We found that the similarity of the retrieved snippets and the ability to copy from them have a significant impact on the results, but the copying capabilities of the 6-layered RETRO were found to be lacking. 
Based on our results, we deem that using small language models with In-context RAG is more sensible than using RETRO for the task of completing lines of code with retrieval from local projects due to the simplicity and broader applicability of In-context RAG.

In summary, users looking to apply our work would do well to consider picking the largest model size that fits within their constraints and then apply retrieval from local projects.
% We confirmed the advantage of RETRO over GPT-2 using single-token metrics on the test set and on individual projects from the test set. 
% We estimated the upper and lower bounds of RETRO based on the retrieval database and examined the impact of the size of the retrieval database. 
% We further confirmed the benefits of RETRO and In-context RAG by completing lines of code in projects from the RepoEval dataset. 
% We demonstrated that In-context RAG significantly improves results for both our smaller and for larger models. 
% We found that the similarity of the retrieved snippets and the ability to copy from them have a significant impact on the results; the copying ability is lacking with the 6-layered RETRO. 
% Due to the simplicity and broader applicability of In-context RAG, we deem that using language models with In-context RAG is more sensible than using RETRO for the task of completing lines of code with retrieval from local projects.

\paragraph{Future work}
RETRO has proven useful for code completion, but several potential improvements would require retraining the model. 
Training could include examples that would improve the ability to copy from retrieved snippets and to ignore irrelevant information. 
For code completion, it would be beneficial to experiment training RETRO with the fill-in-the-middle technique~\citep{fim-donahue-etal-2020-enabling}, which reshapes the context to include code after the cursor position. 
Training RETRO with a longer context could also be enhanced by training on projects instead of individual files.
For better integration with In-context RAG, the retrieved snippets could be appended to the input context during training~\citep{retro++-wang2023shall}. 
Additionally, it would be worthwhile to explore training and using RETRO with different retrieval methods, such as sparse BM-25 search instead of methods based on dense embeddings.

A major advantage of In-context RAG is that it does not require updating the model weights, but doing so could still be beneficial for improving performance either by fine-tuning the model with included retrieved snippets or by fine-tuning the retriever~\citep{dpr-karpukhin-etal-2020-dense, FiD-izacard:hal-03463108}. 
Instead of retrieving snippets based on the Jaccard similarity of tokens, a future area to investigate is incorporating sparse search with the benefits of dense semantic search and check whether hybrid generative and retrieval-based models could further improve performance.
For more efficient execution in code editors, models could be fine-tuned to append to the end of the context instead of the beginning, which would improve the caching of keys and values in self-attention during frequent code editing. 
A next step would also be to explore the impact of combining In-context RAG with the fill-in-the-middle technique, which is becoming increasingly widespread~\citep{wu2024repoformer}.

By training on more data, using appropriate tokenization, and improving the information present in the context with retrieval and fill-in-the-middle, alongside approaches such as weight quantization, we believe that code completion models have great potential for local use in programming tools.

\section*{Acknowledgements}
We thank JetBrains for motivating the problem and providing the computing power for conducting our experiments.

\section*{CRediT authorship contribution statement}
% https://www.elsevier.com/researcher/author/policies-and-guidelines/credit-author-statement
\textbf{Marko Hostnik}: Conceptualization, Methodology, Software, Validation, Formal analysis, Investigation, Writing -- original draft, Writing -- review and editing.
\textbf{Marko Robnik-Šikonja}: Conceptualization, Writing -- original draft, Writing -- review and editing, Supervision, Project administration.

\section*{Funding sources}
This research was funded by the Slovenian Research and Innovation Agency (ARIS) core research programme P6-0411 and project GC-0002. The work was also supported by the EU through ERA Chair grant no. 101186647 (AI4DH) and cofinancing for research innovation projects in support of green transition and digitalisation (project PoVeJMo, no. C3.K8.IB) (Marko Robnik-Šikonja).

\section*{Data availability}
The StarCoder dataset~\citep{li2023starcoder} and the RepoEval~\citep{repocoder-zhang-etal-2023-repocoder} dataset are publicly available.

\appendix

\section{StarCoder dataset analysis}\label{sec:starcoder-dataset-analysis}

\begin{figure*}[htb]
\centering
\includegraphics[width=0.495\linewidth]{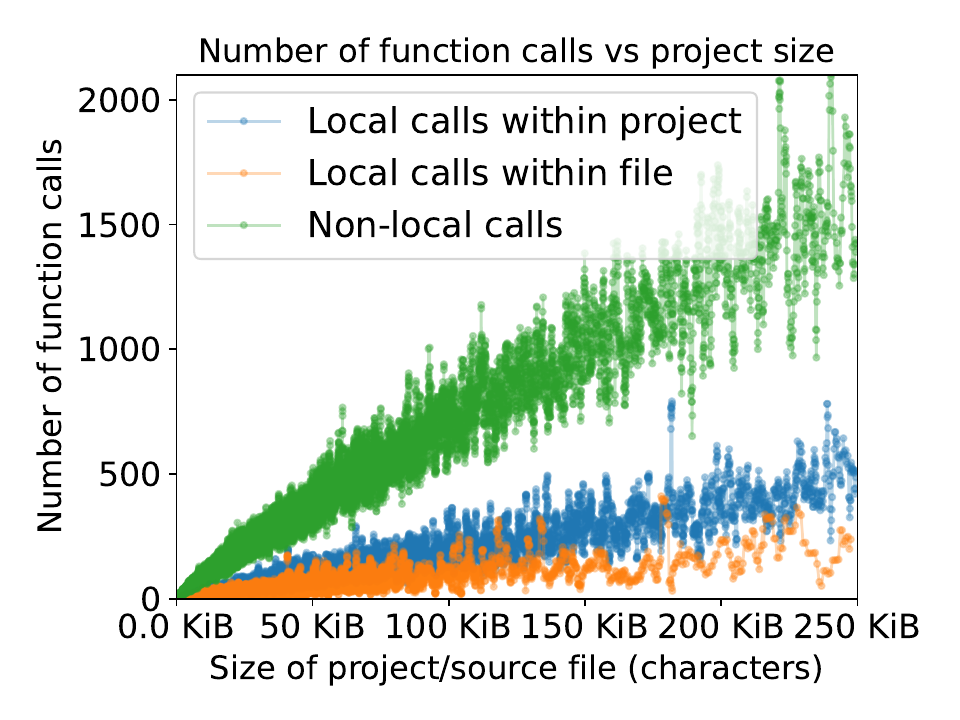}
\includegraphics[width=0.495\linewidth]{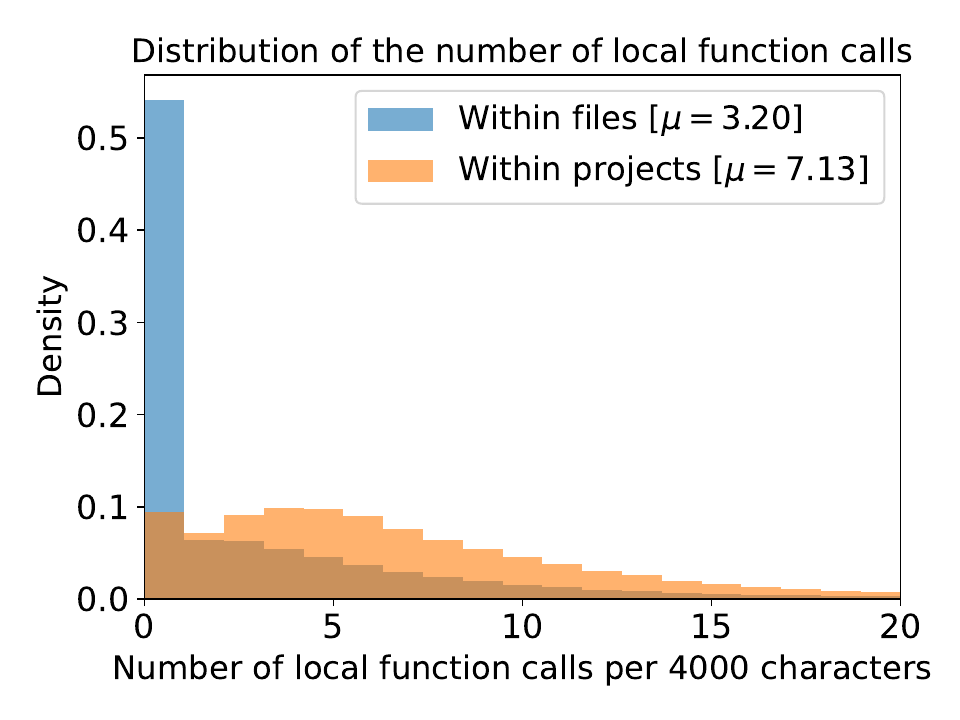}
\caption{
Comparison of references to functions defined within projects in the StarCoder dataset.
Non-local function calls refer to references to functions not defined within the project (e.g., calls to functions from the standard library).
In the right figure, due to varying file lengths, we calculate the average number of calls per \num{4000} characters.
The number of calls increases with project size (left graph), and calls are more frequent when considering the project as a whole compared to individual files (right distribution).
In the right distribution, the number of calls within files stands out near zero, which we attribute to calls to functions defined in other project files.
Within individual files, such calls are counted as non-local.
}
\label{fig:starcoder-calls}
\end{figure*}

In this paper, we examine the benefit of including retrieved similar code snippets from the project into the code completion process.
We hypothesize that finding similar examples can be useful when predicting symbols (e.g., function names, variables, classes) defined within the local project, as such names are not present in the training set.
Therefore, we sample \num{50000} random projects with at least three files from the StarCoder dataset and count the frequency of calls to functions defined within the project by traversing the abstract syntax tree.
In Figure~\ref{fig:starcoder-calls}, we compare the number of calls within individual files to the number of calls within the project as a whole.
The number of calls increases with the size of the project or files, and the number of calls within the project is greater than the number of calls within individual files.
Consequently, we expect that cross-file code retrieval can improve the quality of code completion, especially for larger projects.

\section{RepoEval modification details}\label{sec:repoeval-dataset-details}
The \repoevalline\ and \repoevalapi\ datasets from RepoEval~\citep{repocoder-zhang-etal-2023-repocoder} each contain 1600 examples. %, with 200 examples from each project.
Every example in \repoevalline\ is composed of a \texttt{prompt} input string and a \texttt{ground\_truth} output string.
Our modified dataset \repoevallinerand\ is constructed from \repoevalline\ by expanding each original example's \texttt{prompt} with a random number of characters from \texttt{ground\_truth} and shrinking \texttt{ground\_truth} accordingly.
We are careful that at least one non-whitespace character from \texttt{ground\_truth} is included into the new \texttt{prompt} to avoid expanding \texttt{prompt} only by the leading indentation whitespace.
The construction of \repoevalapirand\ from \repoevalapi\ is similar to that of \repoevallinerand\ from \repoevalline.

\section{Chunk embeddings with BERT}\label{app:bert-codet5p}

We test the impact of using the BERT encoder versus CodeT5+ on a smaller dataset (6.7\% of the total training set) and find that using the larger BERT encoder for chunk embeddings does not result in significant changes in test metrics -- see Table~\ref{tab:bert-codet5}. 
The comparison is not exhaustive due to the smaller training set and shorter training duration. 

\begin{table}[h]
    \centering
    \renewcommand{\arraystretch}{1.1}
    \begin{tabular}{@{}llrrrr@{}}
    \toprule
    Model & Encoder & $\PPL$ & $\Rena$ & $\Rpet$ & $\MRRpet$ \\
    \midrule
    \gptSgpt & / & 2.96 & 77.9 & 89.4 & 82.6 \\
    \retroSgpt & BERT & 2.95 & 78.0 & 89.5 & 82.7 \\
    \retroSgpt & \codetp & 2.95 & 78.0 & 89.4 & 82.7 \\
    \bottomrule
    \end{tabular}
    \caption{
    Comparison of model training on a smaller training dataset using BERT and CodeT5+ for computing chunk embeddings. 
    GPT-2 tokenization is used. 
    % The metrics are defined in Section~\ref{sec:metrics}.
    }
    \label{tab:bert-codet5}
\end{table}

\section{Indexing chunk embeddings}\label{app:indexing}

Due to the large number of chunks in the database, exhaustive searching for the exact $k$-nearest chunks is not feasible, so we resort to \textit{approximate} searching instead. 
As a consequence of approximate search, we seek a compromise between the search accuracy, search speed, index building speed, and index memory usage.
In our implementation, we use the Faiss library~\citep{faissgpu-johnson2019billion} to index the chunk embeddings.\footnote{In Faiss, the used index can be compactly described by the string \texttt{OPQ32\_128,IVF1048576\_HNSW32,PQ32}.}
The search is sped up using IVF indexing~\citep{faissgpu-johnson2019billion} in combination with the HNSW~\citep{hnsw-8594636} algorithm.
Additionally, the vectors are compressed with the PQ~\citep{pq-quantizer-10.1007/978-3-540-88682-2_24} vector codec and the OPQ~\citep{OPQ-Ge_2013_CVPR} transformation.
% We further speed up the search by running it on GPUs.
Table~\ref{tab:tokenization-sizes} summarizes the space requirements for storing the database and retrieval index based on the tokenization method used.

\begin{table}[h]
\centering
\begin{tabular}{@{}lrrrr@{}}
\toprule
Tokenization & \# Chunks & \# Tokens & Tokenized file size & Index size \\
\midrule
GPT-2 & $4.5 \times 10^8$ & $27 \times 10^{9}$ & 53 GiB & 18 GiB \\
StarCoder & $2.8 \times 10^8$ & $18 \times 10^{9}$ & 33 GiB & 12 GiB \\
\bottomrule
\end{tabular}
\caption{Comparison of the retrieval database sizes using two different vocabularies for tokenization.
The total size of all files before tokenization is 62 GiB.}
\label{tab:tokenization-sizes}
\end{table}

% We build the index for searching based on the $L_2$ distance, which is in our case equivalent to searching based on cosine distance due to the properties of the CodeT5+ encoder~\citep{wang-etal-2023-codet5}, which normalizes the output embeddings.
% For $||x|| = ||y|| = 1$, we have:
% \begin{equation*}
% d_{\cos}(x, y) = 1 - \frac{\langle x, y \rangle}{||x|| \cdot ||y||} = 1 - \langle x, y \rangle
% \end{equation*}
% and
% \begin{equation*}
% ||x - y||_2^2 = \langle x - y, x - y \rangle = 2(1 - \langle x, y \rangle) = 2 d{\cos}(x,y).
% \end{equation*}

% In addition to the parameters for building the index, we also need to select parameters that affect the search speed and offer a trade-off between speed and search accuracy. 
% The two parameters that most impact search speed for the built index are the number of clusters visited during the search (for IVF) and the search depth in the graph (for HNSW). 
% We adjust these parameters through experimentation to achieve high accuracy that is acceptably fast for searching.
% We further speed up the search by running it on GPUs.

\paragraph{Indexing effectiveness}

\begin{figure*}[h]
\centering
\includegraphics[width=.495\linewidth]{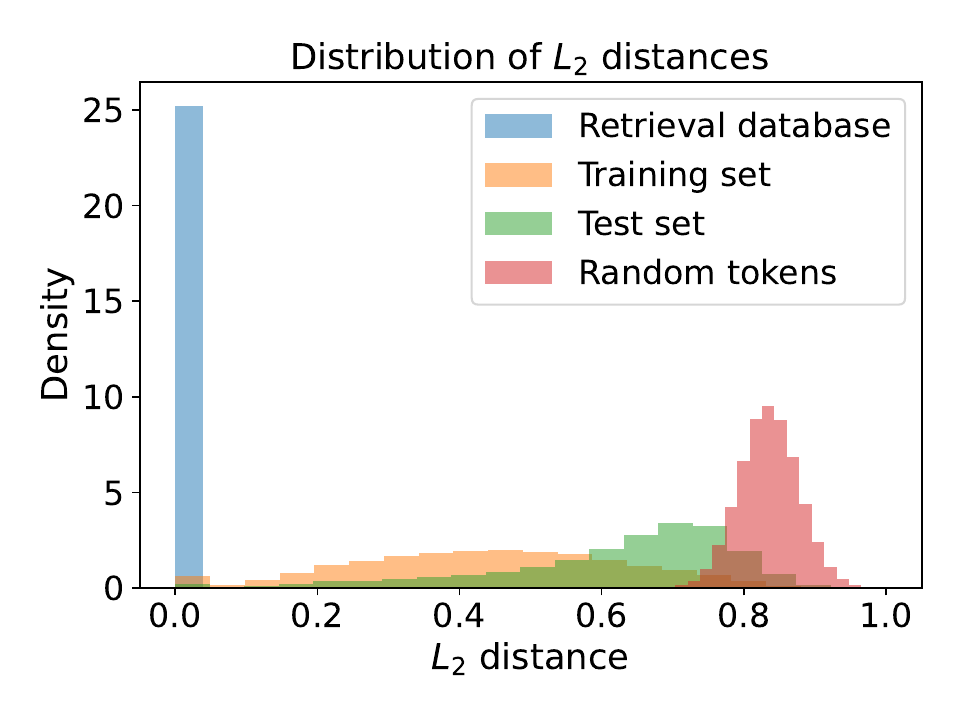}
\includegraphics[width=.495\linewidth]{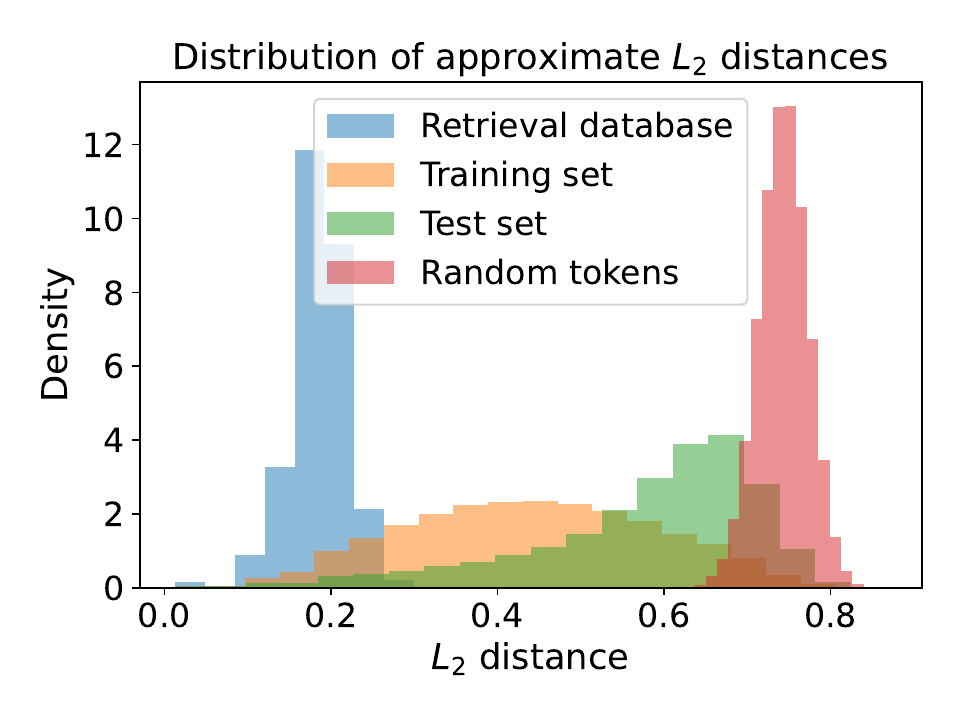}
\caption{
Distribution of $L_2$ distances between chunk embeddings and their retrieved neighbors. 
The left figure shows exact distances, while the right figure shows approximate distances as returned by the Faiss index.}
\label{fig:indexing-distances}
\end{figure*}

To evaluate the effectiveness of approximate nearest chunk retrieval using the index constructed with Faiss, we compare the $L_2$ distance between the embedding of a chunk and the embedding of the nearest chunk from the retrieval database returned by the index.
For comparison, we observe the distributions of $L_2$ distances between the embeddings of retrieved chunks and chunks from the database, training set, test set, and chunks composed of random tokens.
All chunks contain 64 tokens, and we sample \num{100000} chunks from each set.
The distribution results are shown in Figure~\ref{fig:indexing-distances}.
The right side of the figure shows approximate distances as returned by Faiss, which are not exact due to the compression and transformation of embeddings by the indexing.
As expected, retrieving chunks from the database itself is nearly error-free (98\% accuracy), while searching with random chunks returns the largest distances.
Retrieving chunks from the training set results in smaller distances compared to retrieving chunks from the test set, as the database is constructed by partitioning the training set into consecutive non-overlapping chunks.

%%       \citet{<label>} ==> Lamport (1994)
%%       \citep{<label>} ==> (Lamport, 1994)




\bibliographystyle{elsarticle-harv} 
\bibliography{paper}



% \bibitem[Lamport(1994)]{lamport94}
%   \textit{\LaTeX: a document preparation system},
%   Addison Wesley, Massachusetts,

% \end{thebibliography}
\end{document}